\documentclass[letterpaper,titlepage,10pt]{article}

\usepackage{hyperref}

\usepackage{amssymb,amsmath,amsfonts}
\usepackage{epsfig}

\def\clock{{\count0=\time
           \divide\count0 60
           \ifnum\count0<10 0\fi\the\count0
           \multiply\count0 -60 \advance\count0 \time
           :\ifnum\count0<10 0\fi \the\count0
         }}
\newcommand{\timestamp}{{\small\vbox{\hbox{\tt\jobname.tex}
\hbox{\the\day/\the\month/\the\year, \clock}}}}

\newcommand{\CD}{\mathcal{D}}
\newcommand{\CL}{\mathcal{L}}
\newcommand{\CO}{\mathcal{O}}
\newcommand{\CN}{\mathcal{N}}

\newcommand{\CE}{\mathcal{E}}

\newcommand{\Z}{\mathbb{Z}}

\newcommand{\R}{\mathbb{R}}

\newcommand{\nn}{\nonumber}

\newcommand{\spa}{\ , \ \ }

\newcommand{\tr}{\mathop{{\rm Tr}}}

\newcommand{\ads}{\mbox{AdS}}
\newcommand{\gym}{g_{\rm YM}}

\begin{document}

\begin{titlepage}

%\rightline{\vbox{\small\hbox{\tt NORDITA-2008-xx} }}
\ \
 \vskip 2 cm

\centerline{\huge \bf Matching gauge theory and string theory}
\vskip 0.2cm \centerline{\huge \bf in a decoupling limit of AdS/CFT}
\vskip 1.7cm

\centerline{\large {\bf Troels Harmark$\,^{1}$}, {\bf Kristj{\'a}n
R.\ Kristj{\'a}nsson$\,^{2}$} and {\bf Marta Orselli$\,^{1}$} }

\vskip 0.5cm

\begin{center}
\sl $^1$ The Niels Bohr Institute  \\
\sl  Blegdamsvej 17, 2100 Copenhagen \O , Denmark \\
\vskip 0.4cm
\sl $^2$ NORDITA \\
\sl Roslagstullsbacken 23,
10691 Stockholm, Sweden \\
\end{center}
\vskip 0.5cm

\centerline{\small\tt harmark@nbi.dk, kristk@nordita.org,
orselli@nbi.dk}

\vskip 1.5cm

\centerline{\bf Abstract} \vskip 0.2cm \noindent We identify a
regime of the AdS/CFT correspondence in which we can quantitatively
match $\CN=4$ super Yang-Mills (SYM) for small 't Hooft coupling
with weakly coupled type IIB string theory on $\ads_5\times S^5$. We
approach this regime by taking the same decoupling limit on both
sides of the correspondence. On the gauge theory side only the
states in the $SU(2)$ sector survive, and in the planar limit the
Hamiltonian is given by the $XXX_{1/2}$ Heisenberg spin chain. On
the string theory side we show that the decoupling limit corresponds
to a non-relativistic limit. In this limit some of the bosonic modes
and all of the fermionic modes of the string become infinitely heavy
and decouple. We first take the decoupling limit of the string
sigma-model classically. This enables us to identify a
semi-classical regime with semi-classical string states even though
we are in a regime corresponding to small 't Hooft coupling. We
furthermore analyze the quantum corrections that enter in taking the
limit. From this we infer that gauge theory and string theory match,
both in terms of the action and the spectrum, for the leading part
and the first correction away from the semi-classical regime.
Finally we consider the implications for the hitherto unexplained
matching of the one-loop contribution to the energy of certain gauge
theory and string theory states, and we explain how our results give
a firm basis for the matching of the Hagedorn temperature in
hep-th/0608115.

%\vskip 0.5cm \leftline{\timestamp}

\end{titlepage}

\pagestyle{plain} \setcounter{page}{1}

\tableofcontents

%%%%%%%%%%%%%%%%%%%%%%%%%%%%%%%%%%%%%%%%%%%%%%%%%%%%%%%%%%%%%
\section{Introduction}

The duality between gauge theory and string theory plays a major
role in modern theoretical physics. In terms of the AdS/CFT
correspondence \cite{Maldacena:1997re,Gubser:1998bc,Witten:1998qj},
it is responsible for progress in understanding the non-perturbative
behavior of both gauge theory and string theory. It has also led to
insights concerning phenomenologically viable gauge theories (see
for example the review \cite{Mateos:2007ay}).

However, it is difficult to test the AdS/CFT correspondence directly, since the gauge theory and string theory sides usually are not applicable in the
same regime. Indeed, the conventional wisdom is that one needs the
't Hooft coupling $\lambda = \gym^2 N$ to be large, and to be in the
planar limit, in order to see strings in gauge theory, while
perturbative gauge theory calculations only are valid for $\lambda
\ll 1$. In \cite{Harmark:2006di,Harmark:2006ta,Harmark:2007px} a
proposal was put forward for a particular regime of AdS/CFT in which
both gauge theory and string theory are reliable, and hence can be
subject to a detailed match. The regime is
\begin{equation}
\label{regime}
E-J \ll \lambda \ll 1 \spa J \gg 1
\end{equation}
On the gauge theory side, we are considering $SU(N)$ $\CN=4$ SYM on
$\R \times S^3$, and $E$ is the energy of a state measured in units
of the three-sphere radius, while $J= J_1+J_2$ is the sum of two of
the three Cartan generators $J_i$, $i=1,2,3$, of the $SU(4)$
R-symmmetry. On the string theory side, we are considering type IIB
string theory on $\ads_5 \times S^5$, $E$ is the energy of a string
state while $J=J_1+J_2$ is the sum of two of the three Cartan
generators $J_i$, $i=1,2,3$, of the $SO(6)$ symmetry of the
five-sphere, all measured in units of the five-sphere radius.
Moreover, $\lambda=\gym^2 N$ is the 't Hooft coupling of $SU(N)$
$\CN=4$ SYM, which on the string theory side is mapped to
$R^4/(\alpha')^2$, $\sqrt{\alpha'}$ being the string length and $R$
the radius of $\ads_5$ and $S^5$.

The leading part of the dynamics in the regime \eqref{regime}
corresponds to the decoupled theory that one obtains by taking the
following decoupling limit
\cite{Harmark:2006di,Harmark:2006ta,Harmark:2007px}%
\footnote{This decoupling limit was originally conceived as a limit in the Grand Canonical ensemble in which you are close to a critical point with zero temperature and critical chemical potential \cite{Harmark:2006di,Harmark:2006ta}. A closely related limit has been considered in \cite{Harmark:2006ie} corresponding to putting an extra chemical potential in the decoupled theory. In \cite{Harmark:2007px} limits giving other sectors than the $SU(2)$ sector have been found.}
\begin{equation}
\label{limit} \lambda \rightarrow 0 \spa J_i,\, N \ \mbox{fixed}
\spa H \equiv \frac{E-J}{\lambda} \ \mbox{fixed}
\end{equation}
On the gauge theory side, we have in the planar limit $N=\infty$ that $H$ is the Hamiltonian of a ferromagnetic $XXX_{1/2}$ Heisenberg spin chain with the single-trace operators interpreted as states of the spin chain \cite{Minahan:2002ve,Harmark:2006di}. An important ingredient in
this is that only states in the $SU(2)$ sector can survive the
limit. These are the states built only of the two scalars of $\CN=4$
SYM with $J=1$. For all other states of $\CN=4$ SYM it is easy to
see that $E-J$ becomes at least of order one, thus $H$ goes to
infinity in the above limit \eqref{limit}.

For $J \gg 1$ the Landau-Lifshitz sigma-model plus higher derivative
terms gives an effective long wave-length description of the
Heisenberg spin chain \cite{Fradkin}. Using this we observe that we
can find semi-classical states on the gauge theory side, $i.e.$
gauge theory states that have a large value for the sigma-model
action when $J$ is large.%
\footnote{We find semi-classical string states when $\lambda/(E-J)
\sim J$ with $\lambda \ll 1$ and $J \gg 1$, thus we are in the
regime~\eqref{regime}.}
 This could seem surprising in that we are
in weakly coupled gauge theory $\lambda \ll 1$, $i.e.$ it
contradicts the standard lore that one should only find
semi-classical string states for $\lambda \gg 1$. This observation
motivates us to show that the regime \eqref{regime} can be a
semi-classically valid regime for strings on $\ads_5\times S^5$ even
though the effective string tension is small $R^2 / \alpha' \ll 1$,
which normally would mean that we are deep into a quantum string
regime.

On the string side the limit \eqref{limit} can be written as
\begin{equation}
\label{strlimit} \frac{R^2}{\alpha'} \rightarrow 0 \spa J \ \mbox{fixed}
\spa H \equiv \frac{(\alpha')^2}{R^4} ( E-J) \ \mbox{fixed} \spa
\tilde{g}_s \equiv g_s \frac{(\alpha')^2}{R^4} \ \mbox{fixed}
\end{equation}
This limit involves taking the effective string tension $R^2/\alpha'$ to zero, again suggesting that we are deep into a quantum string regime. However, we find that when taking the limit \eqref{strlimit} on the sigma-model for $\ads_5\times S^5$ the action for the surviving string modes remains finite and is moreover large when $J$ is large. $I.e.$ writing schematically the bosonic sigma-model for $\ads_5\times S^5$ as
\begin{equation}
I = - \frac{R^2}{4\pi \alpha'} \int d^2\sigma G_{\mu \nu} \partial^\alpha X^\mu \partial_\alpha X^\nu
\end{equation}
we find that the action remains finite in the limit \eqref{strlimit}
due to the fact that $\int d^2\sigma G_{\mu \nu} \partial^\alpha
X^\mu \partial_\alpha X^\nu$ scales like $J \alpha' / R^2$ in the
limit \eqref{strlimit}, thus making $J$ the effective string tension
in the regime \eqref{regime}. The regime \eqref{regime} is therefore
a new semi-classical regime of type IIB string theory on $\ads_5
\times S^5$. This is in agreement with the gauge theory side where
we also find semi-classical string states in the regime
\eqref{regime}.

Taking the limit \eqref{strlimit} on the level of the classical string theory sigma-model we end up with the Landau-Lifshitz sigma-model. This resembles a similar limit of the classical bosonic sigma-model on $\R \times S^3$ considered by Kruczenski \cite{Kruczenski:2003gt}.%
\footnote{The way we take the limit \eqref{strlimit} of the
classical sigma-model on $\R\times S^3$ resembles closely the limit
of Kruczenski \cite{Kruczenski:2003gt}. However, the limit is not
the same as the one considered by Kruczenski. The most important
difference is that we do not assume we are in the semi-classical
regime $R^2/\alpha' \gg 1$ in our limit. This is connected to the
fact that we consider closely how the quantum effects come into play
in our limit. It is also important to remark that the way we take
our limit of the sigma-model is completely determined from the limit
\eqref{limit}.} We consider subsequently the possible quantum
corrections to the string theory sigma-model that can contribute in
the limit \eqref{strlimit}. One reason that our analysis holds is
due to the exactness of the supersymmetric string action on
$\ads_5\times S^5$
\cite{Kallosh:1998qs,Metsaev:1998it,Kallosh:1998zx,Frolov:2006cc}.
Another important aspect is the decoupling of six transverse bosonic
fields, plus all the fermionic fields, which plays a crucial role.
We argue that these modes become infinitely heavy and thus decouple
in the limit \eqref{strlimit} and through integrating them out they
can only show up as higher-derivative terms for the surviving modes.
In addition, we argue that zero-mode quantum effects for the
decoupled modes are absent since we are close to $E=J$ which
corresponds to half-BPS supersymmetric states.

By analyzing the classical sigma-model and the quantum effects, we
conclude that the limit \eqref{strlimit} gives the Landau-Lifshitz
sigma-model up to $1/J^2$ corrections where the quantum effects can
set in. The quantum effects enters as higher derivative terms coming
from integrating out the decoupled modes. We can therefore match the
effective sigma-model action for the strings, up to order $1/J^2$
corrections, to the sigma-model action obtained on the gauge theory
side by considering large $J$. This enables us furthermore to show
that not only we can match the leading order energy of
semi-classical states but also the energy of quantum string states,
up to $1/J^2$ corrections.

It might seem like we have found a string/gauge-theory duality which is a weak-weak duality. However, this is not the case. Instead, what we see on the gauge theory side is that the effective coupling is not $\lambda$ but rather $\lambda / (E-J)$ when taking the limit \eqref{limit}. Thus, for the gauge theory, \eqref{regime} is really a strong coupling regime since $\lambda / (E-J) \gg 1$. However, differently from usual, we have complete control over this regime by only knowing the one-loop contribution to the anomalous dimension of gauge theory operators.%
\footnote{As pointed out to us by Erik Verlinde, our limit has
similarities with the 't Hooft limit where $N \rightarrow \infty$
with $\lambda = \gym^2 N$ fixed. Here $\gym^2$ is sent to zero in
the limit but $\lambda \gg 1$ is still a strong coupling regime.
Moreover, in the 't Hooft limit you access a simpler strong coupling
regime than in the finite $N$ theory (since you only have planar
diagrams) which is somewhat analogous to the situation in our limit
where we have full control over the strong coupling regime.}
Moreover, the identification of the one-loop dilatation operator as
a spin chain is crucial for understanding the spectrum of the gauge
theory side in the regime \eqref{regime}. Therefore, in this sense
it should not be surprising that the regime \eqref{regime} is under
control in weakly coupled string theory, in that it corresponds to a
particular kind of strong coupling regime of the gauge theory side.

In previous work on matching gauge theory and string theory in the AdS/CFT correspondence the starting point is that one should connect the weakly coupled gauge theory regime $\lambda \ll 1$ with the semi-classical string theory regime $\lambda \gg 1$. In particular, for gauge theory and string theory states in the $SU(2)$ sector one can on both sides of the AdS/CFT correspondence make the expansion in $\lambda' = \lambda /J^2$ of the energy as follows
\begin{equation}
E-J = \lambda' E_1 + \lambda'^2 E_2 +  \cdots
\end{equation}
since $J$ is large on both sides. In particular, it has been
observed that at order $\lambda'$, $i.e.$ the one-loop contribution
on the gauge theory side, you find the same energy from gauge theory
and string theory up to $1/J^2$ corrections
\cite{Minahan:2002ve,Beisert:2003tq,Callan:2003xr,Callan:2004uv,Frolov:2003qc,Beisert:2005mq,Astolfi:2008yw},
even though you compute it in two different regimes of the AdS/CFT
correspondence. This matching of the energies begs for an
explanation. Using our results we are able to provide this
explanation by giving a simple argument for why one should obtain
the same result for the $\lambda'$ contribution for string theory in
the regime $\lambda \gg 1$ as in the regime \eqref{regime}. This
relies on our result that the effective sigma-model for type IIB
string theory on $\ads_5\times S^5$ in the limit \eqref{strlimit} is
given by the Landau-Lifshitz sigma-model up to $1/J^2$ corrections.

It is important to note that in the limit \eqref{limit} we have
$\lambda' = \lambda / J^2 \rightarrow 0$ hence this corresponds to
taking a large volume limit with respect to the wrapping
interactions for the spin chain description of $\CN=4$ SYM \cite{Beisert:2004hm,Ambjorn:2005wa},
$i.e.$ it is a limit in which wrapping effects are suppressed and
the phase factor in the S-matrix description of the asymptotic Bethe
equations for $\CN=4$ SYM is trivial \cite{Arutyunov:2004vx,Beisert:2005tm,Beisert:2006ez}. Thus, our results show
that one can match gauge theory and string theory in the AdS/CFT
correspondence in this regime. This is consistent with the fact that
the conjectured Bethe equations for quantum strings
\cite{Arutyunov:2004vx} become the Bethe equations for the
Heisenberg sigma-model in the limit \eqref{limit}. Thus, the results
of this paper provide an argument for why this should be the case.

We discuss furthermore the physical interpretation of our decoupling
limit \eqref{limit}. On the string theory side, we show that the
limit \eqref{strlimit} (or equivalently \eqref{limit}) in fact is a
non-relativistic limit for type IIB string theory on $\ads_5\times
S^5$. $I.e.$ it is a low energy limit and a limit of slow velocities
for the strings. Moreover, we show that the decoupling of certain
modes of the strings corresponds to going from a relativistic field
theory, where we have an anti-particle for each particle, to a
Galilean field theory. This is furthermore connected to the fact
that we obtain a space-space non-commutative theory in the limit
\eqref{strlimit}. We explain that this is because the effective
sigma-model should describe a one-dimensional spin chain, hence the
two spatial directions become the two directions in a phase space
for a single spatial direction.

We consider briefly the interplay between the decoupling limit
\eqref{limit} and the Penrose limit of \cite{Bertolini:2002nr},
which is a geometric limit of the $\ads_5 \times S^5$ background
giving the maximally supersymmetric pp-wave background of
\cite{Blau:2001ne}, in the coordinate system with a flat direction
\cite{Michelson:2002wa,Bertolini:2002nr}. We explain that we can
consider the two limits in different successions and that based on
the results of this paper one finds the same limiting theory, which
is a free theory with Galilean symmetry, regardless of the
succession of the limits.

Finally, we consider the implications of the results of this paper
for the matching of the Hagedorn temperature in
\cite{Harmark:2006ta} (see also \cite{Harmark:2006ie}). We explain
that the results of this paper puts the matching of the Hagedorn
temperature on a firm basis since they show that one can match the
leading order spectra of gauge theory and string theory in the limit
\eqref{limit} for large $J$. We can conclude from this that the
Hagedorn temperature is the first example of a quantity not
protected by supersymmetry that has been interpolated successfully
from the weakly coupled gauge theory to the semi-classical string
theory regime.

%%%%%%%%%%%%%%%%%%%%%%%%%%%%%%%%%%%%%%%%%%%%%%%%%%%%%%%%%%%%%
\section{Decoupling limit of planar $\CN = 4$ SYM on $\R \times S^3$}
\label{sec:decgauge}

In this section we review briefly the decoupling limit of $\CN=4$
SYM on $\R \times S^3$
\cite{Harmark:2006di,Harmark:2006ta,Harmark:2007px} giving a decoupled theory with an $SU(2)$ symmetry.%
\footnote{There are altogether 12 non-trivial decoupled theories
\cite{Harmark:2007px}. These 12 theories correspond to twelve
different classes of limits of $\CN=4$ SYM on $\R \times S^3$. In
the grand canonical ensemble they correspond to being close to
either of the twelve different critical points.} We consider it here
in terms of the charges and the energy/scaling dimension, $i.e.$ in
the microcanonical ensemble. We review in particular that the
decoupled theory in the planar limit corresponds to the
ferromagnetic Heisenberg $XXX_{1/2}$ spin chain, and that one can
take a continuum limit in which we can approximate the Heisenberg
spin chain by a sigma-model.

Note that we discuss in Section \ref{sec:hag} and in the conclusions
in Section \ref{sec:concl} why $\lambda/(E-J) \gg 1$ can be seen as
a strong coupling regime.

\subsection{Review of decoupling limit}

We review here briefly the $SU(2)$ decoupling limit of
\cite{Harmark:2006di,Harmark:2006ta,Harmark:2007px} on the gauge
theory side of the AdS/CFT correspondence. We are considering
$SU(N)$ $\CN=4$ SYM on $\R \times S^3$. Since we take the large $N$
limit below we introduce the 't Hooft coupling $\lambda = \gym^2 N$.
We denote the three R-charges for the $SO(6) \simeq SU(4)$
R-symmetry as $J_1,J_2,J_3$. We employ in the following the
state/operator correspondence relating a state of $\CN=4$ SYM on $\R
\times S^3$ of energy $E$ to an operator of $\CN=4$ SYM on $\R^4$ of
scaling dimension $D = E$, $i.e.$ we set the radius of the $S^3$ to
one. Due to the compactification on $S^3$ the states are restricted
to be singlets of $SU(N)$ which on the operator side restricts us to
the class of operators consisting of linear combinations of
multi-trace operators.

\subsubsection*{The $SU(2)$ decoupling limit of $\CN=4$ SYM on $\R
\times S^3$}

We consider the following decoupling limit of $SU(N)$ $\CN=4$
SYM on $\R \times S^3$
\cite{Harmark:2006di,Harmark:2006ta,Harmark:2007px}
\begin{equation}
\label{gaugelimit} \lambda \rightarrow 0 \spa J_i,\, N \
\mbox{fixed} \spa H \equiv \frac{E-J}{\lambda}  \ \mbox{fixed}
\end{equation}
with $J \equiv J_1 + J_2$ and $\lambda = \gym^2 N$. Note in
particular that $J$ is fixed in the limit. In terms of operators we have
that $H=(D-J)/\lambda$. The scaling dimension is found by
diagonalizing the dilatation operator $D$ \cite{Beisert:2002ff,
Beisert:2003tq,Beisert:2004ry}. At weak 't Hooft
coupling we expand $D$ as
\begin{equation}
D = D_0 + \lambda D_2 + \lambda^{\frac{3}{2}} D_3 + \lambda^2 D_4 + \cdots
\end{equation}
Here $D_0$ is the bare scaling dimension and $D_2$ is the one-loop
part of the dilatation operator (see \cite{Beisert:2003jj} for a
complete expression). Taking now the limit \eqref{gaugelimit} we see
that since $D_0-J$ is a integer or half-integer we have  that
$(D-J)/\lambda$ goes to infinity in the limit unless $D_0=J$. Thus,
all operators with $D_0 > J $ decouple in the limit. Note here that
any operator in $\CN=4$ SYM obeys the bound $D_0 \geq J $. The class
of operators saturating the bound, $i.e.$ that has $D_0=J$,
corresponds to the so-called $SU(2)$ sector of $\CN=4$ SYM
consisting of all possible operators that one can make from linear
combinations of multi-trace operators built from the single-trace
operators of the form
\begin{equation}
\label{sing}
\tr ( A_1 A_2 \cdots A_J ) \spa A_i \in \{ Z,X \}
\end{equation}
Here $Z$ and $X$ are two of the three complex scalars of $\CN=4$ SYM with R-charges $(J_1,J_2,J_3)=(1,0,0)$ for $Z$ and $(J_1,J_2,J_3)=(0,1,0)$ for $X$. Since $J=J_1+J_2$ we see that the total number of $Z$'s and $X$'s add up to $J$ for any operator.%
\footnote{Note that we keep all three R-charges fixed in the limit \eqref{gaugelimit} thus we should take the limit with $J_3=0$ since if $J_3 \neq 0$ we decouple all operators.}

For states in the $SU(2)$ sector we see now that from
$H=(D-J)/\lambda$ we get an effective Hamiltonian
\begin{equation}
H = D_2
\end{equation}
in the limit \eqref{gaugelimit}. This is a Hamiltonian in the sense that we get the energies/scaling dimensions of the surviving states/operators by diagonalizing $H$. We have explicitly \cite{Beisert:2002bb,Beisert:2003tq}
\begin{equation}
H = - \frac{\lambda}{8 \pi^2 N} \tr [X,Z][\bar{X},\bar{Z}]
\end{equation}
where $\bar{X} = \delta / \delta X$ and $\bar{Z} = \delta / \delta Z$. This Hamiltonian gives the spectrum of our decoupled theory.

\subsubsection*{Planar limit corresponds to ferromagnetic Heisenberg
chain}

Considering now the planar limit $N=\infty$ we can employ large $N$
factorization and get the scaling dimension of any operator from
knowing the scaling dimension of single-trace operators. Also, by
the same token the mixing between single-trace operators and
multi-trace operators goes to zero. Thus, we can get the whole
spectrum by just focusing on the single-trace operators. Considering
now the single-trace operators in the $SU(2)$, we have that a
single-trace operator \eqref{sing} can be interpreted as a state in
a spin $1/2$ spin chain \cite{Minahan:2002ve} with the letters $Z$ and $X$
being the spin up and spin down state.

In detail we choose the spin as $S_z = (J_1-J_2)/2$ which means that
$S_z=1/2$ for $Z$ and $S_z=-1/2$ for $X$. Thus, for each site on the
spin chain we have a two-dimensional vector-space spanned by the
spin-up and spin-down states. On this two-dimensional space we can
define the spin vector $\vec{S}_i$ for site number $i$ as $\frac{1}{2} \vec{\sigma}$ acting on the state of the $i$'th site,
where $\vec{\sigma}$ are the Pauli matrices. In this way we see that $S_z = \sigma_z /2$ which is consistent with the above definition of $S_z$. The Hamiltonian $H$, as defined by the limit \eqref{limit}, is then given by
\begin{equation}
\label{gaugeham} H = \frac{1}{4\pi^2} \sum_{i=1}^J \left(
\frac{1}{4} - \vec{S}_i \cdot \vec{S}_{i+1} \right)
\end{equation}
This is the Hamiltonian for the ferromagnetic $XXX_{1/2}$ Heisenberg
spin chain with zero magnetic field \cite{Minahan:2002ve}. Here $J$
is the length of the spin chain and $S_z = (J_1-J_2)/2$ is the total
spin. Thus, in conclusion, this is the decoupled theory that one
gets from taking the decoupling limit \eqref{gaugelimit} of planar
$\CN=4$ SYM on $\R \times S^3$.

Since the ferromagnetic $XXX_{1/2}$ Heisenberg spin chain is an
integrable system we can write down the following equations that in
principle determines the full spectrum of $H$ %
\footnote{See \cite{Harmark:2006ta} for a construction of the Bethe ansatz
that takes into account that the total spin $S_z = (J_1-J_2)/2$ is
fixed in our decoupled theory.}
\begin{equation}
\label{exact1} H = \frac{1}{2\pi^2} \sum_{i=1}^M \sin^2 \left(
\frac{p_i}{2} \right)
\end{equation}
\begin{equation}
\label{exact2} e^{ip_k J } = \prod_{j=1,j \neq k}^M S(p_k,p_j) \spa
S(p_k,p_j) = -
\frac{1+e^{i(p_k+p_j)}-2e^{ip_k}}{1+e^{i(p_k+p_j)}-2e^{ip_j}} \spa
\sum_{i=1}^M p_i = 0
\end{equation}
These equations are the dispersion relation for $H$, the Bethe
equations along with the S-matrix, and a zero total momentum
condition due to the cyclicity of the trace. We have introduced here
$M$ momenta $p_i$ corresponding to $M$ magnons which are
pseudoparticles propagating on the chain.

For $J$ large, we can consider the low energy part of the spectrum
$H \ll 1$. This corresponds to having the momenta of the magnons of
order $1/J$ to leading order. One then finds from
\eqref{exact1}-\eqref{exact2} the following leading order low energy
spectrum of $H$
\begin{equation}
\label{specgauge2} H = \frac{1}{2 J^2} \sum_{n \neq 0} \left( 1 +
\frac{2}{J} \right) n^2 M_n  + \CO(1/J^2) \spa \sum_{n\neq 0} n M_n
= 0
\end{equation}
where $M_n$ is the number operator for the integer level $n$ with $n
\neq 0$. Note that this spectrum is only true for states built from
magnons with different momenta $p_i$, $i.e.$ it is not true for
bound states. We ignore this subtlety for simplicity of
presentation.

We see from the spectrum \eqref{specgauge2} that for large $J$ the
Hamiltonian $H$ goes like $1/J^2$ for the low energy excitations. It
is therefore natural in this regime to introduce the rescaled
Hamiltonian $\tilde{H}$
\begin{equation}
\label{tildeH} \tilde{H} = J^2 H = \frac{J^2}{\lambda} (E-J) =
\frac{J^2}{4\pi^2} \sum_{i=1}^J \left( \frac{1}{4} - \vec{S}_i \cdot
\vec{S}_{i+1} \right)
\end{equation}
such that the spectrum \eqref{specgauge2} is
\begin{equation}
\label{specgauge3} \tilde{H} = \frac{1}{2} \sum_{n \neq 0} \left( 1
+ \frac{2}{J} \right) n^2 M_n  + \CO(1/J^2) \spa \sum_{n\neq 0} n
M_n = 0
\end{equation}
We see that focusing on the low energy part of the spectrum of $H$
in the large $J$ limit corresponds to considering the part of the
spectrum of $\tilde{H}$ which is of order one.

Note that $\tilde{H}\sim 1$ means that $\lambda/(E-J) \sim J^2$.
Thus, this falls within the regime \eqref{regime}. Therefore we can
conclude that we find quantum string states in planar $\CN=4$ SYM on
$\R\times S^3$ in the regime \eqref{regime}.

%%%%%%%%%%%%%%%%%%%%%%%%%%%%%%%%%%%%%%%%%%%%%%%%%%%%%%%%%%%%%%%%%
\subsection{Effective sigma-model description in continuum limit}
\label{sec:smgauge}

In this section we review that the ferromagnetic Heisenberg spin
chain, that we obtain from our decoupling limit \eqref{limit} of
planar $\CN=4$ SYM on $\R\times S^3$, has a sigma-model description
for large $J$ \cite{Fradkin}. We review this briefly in order to
compare to what happens on the string side in the decoupling limit
\eqref{limit} in Section \ref{sec:decstrings}.

In the context of AdS/CFT, the sigma-model limit of the Heisenberg
spin chain was first used in \cite{Kruczenski:2003gt}. We comment below,
and also in the rest of this paper, in what sense our approach is
different from that of \cite{Kruczenski:2003gt}.

To obtain a sigma-model description of the Heisenberg spin chain,
one begins by introducing a coherent state $| \vec{n} \rangle$ for
each site of the spin chain such that
\begin{equation}
\langle \vec{n} | \vec{\sigma} | \vec{n} \rangle = \vec{n}
\end{equation}
where $\vec{\sigma}$ are the two by two Pauli matrices. Here
$\vec{n}$ is a unit vector pointing to a point on the two-sphere
parameterized as
\begin{equation}
\label{vecn} \vec{n} = (\cos \theta \cos \varphi, \cos \theta \sin
\varphi, \sin \theta )
\end{equation}
One then proceeds to first write up the one-spin partition function,
ignoring the interaction between different spins. This can be done
using the usual derivation of the path-integral in quantum
mechanics. One can then use this to write up the partition function
for the full spin chain, now including the interaction Hamiltonian
\eqref{tildeH}. Altogether, this gives the partition function and
action \cite{Fradkin}
\begin{equation}
\label{ex_act} Z = \int \CD \vec{n} e^{i I [\vec{n}]} \spa
I[\vec{n}] = \sum_{k=1}^J \int d\tilde{t} \left[ \vec{C} (\vec{n}_k)
\cdot
\partial_{\tilde{t}} \vec{n}_k - \frac{J^2}{32\pi^2} ( \vec{n}_{k+1}
- \vec{n}_{k} )^2 \right]
\end{equation}
where
\begin{equation}
\vec{C} (\vec{n}) \cdot \partial_{\tilde{t}} \vec{n} = - \frac{1}{2}
\int_0^1 d\xi \epsilon_{ijk} n_i \partial_\xi n_j
\partial_{\tilde{t}} n_k = \frac{1}{2}
\sin \theta
\partial_{\tilde{t}} \varphi
\end{equation}
is a Wess-Zumino type term where $\vec{C}(\vec{n})$ is proportional
to the area spanned between the trajectory and the north pole of the
two-sphere \cite{Fradkin}. The action \eqref{ex_act} provides an
equivalent description of the Heisenberg spin chain. Notice that
this action is describing a one-dimensional lattice of $J$ spins,
$i.e.$ $\vec{n}_k$ is the spin on the $k$'th site of the lattice.

It is important to note that in deriving the action \eqref{ex_act}
we have used a time $\tilde{t}$ corresponding to $\tilde{H}$, $i.e.$
$\tilde{H} = J^2(E-J)/\lambda = i \partial_{\tilde{t}}$. This is
because we are interested in the low energy dynamics for $J$ large,
hence the regime in which $\tilde{H}$ is of order one.

Taking now the limit $J \rightarrow \infty$, we can approximate the
one-dimensional lattice by a continuous variable. Denoting this
variable $\sigma$ we are considering the field $\vec{n}
(\tilde{t},\sigma)$. Therefore, imposing that $\sigma$ has period
$2\pi$, we should map the $k$'th site to $\sigma=2\pi k/J$, $i.e.$
$\vec{n}_k (\tilde{t})$ is mapped to $\vec{n} (\tilde{t}, \sigma)$.
Correspondingly we map the sum $\sum_{k=1}^J$ to the integral
$\frac{J}{2\pi} \int_{0}^{2\pi} d\sigma$. We furthermore use that
\begin{equation}
\vec{n}_{k+1} - \vec{n}_{k} = \exp \left( \frac{2\pi}{J}
\partial_\sigma \right) \vec{n} - \vec{n}
\end{equation}
This gives the action
\begin{equation}
\label{contact} I[\vec{n}] = \frac{J}{2\pi} \int d\tilde{t}
\int_0^{2\pi} d\sigma \left[ \vec{C} (\vec{n}) \cdot
\partial_{\tilde{t}} \vec{n} + \frac{J^2}{8\pi^2} \vec{n} \cdot \sinh^2 \left(
\frac{\pi}{J} \partial_\sigma \right) \vec{n} \right]
\end{equation}
We see that this can be considered to be a sigma-model on a
continuous world-sheet parameterized by $\tilde{t}$ and $\sigma$,
with the target space given by $S^2 \simeq SU(2) / U(1)$. The first
term in \eqref{contact} is a kinetic term, while the second one is a
potential term. The second term is responsible for the dispersion
relation \eqref{exact1}. This can be seen from the fact that the
momenta $p$ of an impurity is mapped to $-i (2\pi
/J) \partial_\sigma$. We see that the discreteness of the Heisenberg
spin chain manifests itself as a infinite sum over higher derivative
terms in the continuum action \eqref{contact}.

Expanding now the action \eqref{contact} in powers of $1/J$ we have
\begin{equation}
I[\vec{n}] = \frac{J}{2\pi} \int d\tilde{t} \int_0^{2\pi} d\sigma
\left[ \vec{C} (\vec{n}) \cdot  \dot{\vec{n}} - \frac{1}{8}
(\vec{n}')^2 + \frac{\pi^2}{24 J^2} (\vec{n}'')^2 + \CO(J^{-4})
\right]
\end{equation}
where we introduced a dot (prime) as the derivative with respect to
$\tilde{t}$ ($\sigma$). The leading part of the action
\eqref{contact} in the thermodynamic limit $J \rightarrow \infty$ is
therefore
\begin{equation}
\label{leadint} I[ \theta, \varphi ] \simeq I_{\rm LL}
[\theta,\varphi] \equiv \frac{J}{4\pi} \int d\tilde{t} \int_0^{2\pi}
d\sigma \left[ \sin \theta \dot{\varphi} - \frac{1}{4} \left[
(\theta')^2 + \cos^2 \theta (\varphi')^2 \right]
 \right]
\end{equation}
here written in terms of the parametrization \eqref{vecn}. We see
that to leading order in the thermodynamic limit the Heisenberg spin
chain is well-described by the Landau-Lifshitz model with action
$I_{\rm LL} [\theta,\varphi]$.

Finally, we note that the constraint of zero total momentum
$\sum_{i=1}^M p_i=0$ in \eqref{exact2} takes the following form for
the sigma-model $\int_0^{2\pi} d\sigma \sin \theta \varphi' = 0$ in
terms of the parametrization \eqref{vecn}.

\subsubsection*{Getting the spectrum from the sigma-model}

We now explain briefly how to get the spectrum \eqref{specgauge3}
from the sigma-model. We begin by considering a limit of the action
\eqref{contact}, here dubbed the free limit, in which the
sigma-model reduces to a free theory with Galilean symmetry. The
free limit is a large $J$ limit in which we zoom in to a point on
the equator of the two-sphere. More specifically, it is a limit in
which we zoom in near the point $(\theta,\varphi)=(0,0)$ such that the
two-sphere metric $d\Omega_2^2 \simeq d\theta^2 + d\varphi^2$, which
means that the geometry near this point is 2D Euclidean space. We
take the free limit by defining the rescaled coordinates
\begin{equation}
\label{xy} x = \sqrt{J} \varphi \spa y = \sqrt{J} \theta
\end{equation}
which we keep fixed as $J\rightarrow \infty$. Taking now the
$J\rightarrow \infty$ limit of the action \eqref{leadint} we get the
following action
\begin{equation}
\label{galact} I = \frac{1}{4\pi} \int d\tilde{t} \int_0^{2\pi}
d\sigma \left[ y \dot{x} - \frac{(x')^2 + (y')^2}{4} \right]
\end{equation}
This action can easily be quantized. Define $z = x+iy$. Then we can
write the EOMs as $\dot{z} = \frac{i}{2} z''$. Using the EOMs plus
the periodicity of $\sigma$ we see that the general expansion of
$z(\tilde{t},\sigma)$ is
\begin{equation}
\label{zz} z(\tilde{t},\sigma) = 2  \sum_{n\in \Z} a_n e^{-i
\frac{n^2}{2} \tilde{t} + in\sigma}
\end{equation}
To quantize the theory, we note that from the action \eqref{galact}
we have that the conjugate momentum to $x$ is $p_x = y/(4\pi)$. The
canonical commutation relation is
\begin{equation}
[x(\tilde{t},\sigma), p_x (\tilde{t},\sigma')] = i
\delta(\sigma-\sigma')
\end{equation}
Using this with \eqref{zz} we see that the $a_n$'s becomes lowering
operators with the canonical commutation relation
\begin{equation}
[a_n,a_k^\dagger] = \delta_{nk}
\end{equation}
We see from this that $z(\tilde{t},\sigma)$ only contains lowering
operators. This means that we do not have an anti-particle part of
$z(\tilde{t},\sigma)$. This fits with the fact that we have a
non-relativistic dispersion relation $H \propto p_\sigma^2$
suggesting that we do not have an antiparticle propagating backwards
in time like in relativistic field theory.%
\footnote{The non-relativistic nature of the Landau-Lifshitz
sigma-model is also considered in \cite{Klose:2006dd}.} We remark
furthermore that since we have that $y = 4\pi p_x$ the two
transverse dimensions that we started with have become the
two-dimensional phase space for the one dimension $x$. Thus we see a
reduction from two space-like dimensions to just one space-like
dimension. This is connected to the non-relativistic nature of the
action \eqref{galact} since normally two spatial directions in a
sigma-model would give rise the double number of raising and
lowering operators as we found above. We consider further the
non-relativistic nature of the Heisenberg model in Section
\ref{sec:nonrel}.

From the action \eqref{galact} we see that the Hamiltonian is
\begin{equation}
\tilde{H} = \int_0^{2\pi} d\sigma \left[ \frac{(x')^2}{16\pi} + \pi
(p_x')^2 \right]
\end{equation}
Inserting now \eqref{zz} in this, we get the spectrum
\begin{equation}
\label{string0} \tilde{H} = \frac{1}{2} \sum_{n \in \Z} n^2 M_n \spa
\sum_{n \in \Z} n M_n = 0
\end{equation}
with the number operator being $M_n = a_n^\dagger a_n $. This
spectrum matches the leading order part of the spectrum
\eqref{specgauge3}. The second equation is the level-matching
condition derived by imposing the
vanishing of the total world-sheet momentum.%
\footnote{This is derived from $\int_0^{2\pi} d\sigma  y x'  = 0$.}

%%%%%%%%%

One can also find easily the $1/J$ correction to the leading
spectrum \eqref{specgauge3} from the sigma-model (see for example
\cite{Minahan:2005mx}). We begin using the coordinate $x$ as defined in
\eqref{xy}. The conjugate momentum $p_x$ for the full action
\eqref{thesigmod} is
\begin{equation}
\label{px} p_x = \frac{\sqrt{J}}{4\pi} \sin \theta
\end{equation}
Inserting this in the action \eqref{thesigmod}, we can easily derive
the Hamiltonian
\begin{equation}
\label{Hcorr} \tilde{H} = \int_0^{2\pi} d\sigma \left[ \left( 1 -
\frac{16\pi^2}{J} p_x^2 \right) \frac{(x')^2}{16\pi} + \left( 1+\frac{16\pi^2}{J} p_x^2 \right) \pi
(p_x')^2 + \CO ( J^{-2} ) \right]
\end{equation}
To find the $1/J$ corrections one can use ordinary quantum
mechanical perturbation theory and plug in the zeroth order $x$ and
$p_x$, as found from \eqref{zz}, into the Hamiltonian \eqref{Hcorr}.
Doing this, one obtains precisely the corrected spectrum
\eqref{specgauge3}.%
\footnote{Note again that the spectrum \eqref{specgauge3} only
describes non-bound-states, $i.e.$ states build from raising
operators with different levels. The leading part of the spectrum
\eqref{string0} is instead true for all states.}

\subsubsection*{Semi-classical states in decoupled theory}

The spectrum \eqref{specgauge3} which has $\tilde{H}$ of order one,
corresponds to considering a finite number of impurities for the
Bethe equations \eqref{exact1}-\eqref{exact2}. If we instead
consider a number of impurities $M$ such that $M/J$ is of order one,
we get that the energy $\tilde{H}$ of such states is of order $J$.
Such states are semi-classical since when considering large quantum
numbers we can approximate the quantum physics with classical
physics.

From the sigma-model point of view we have a natural classical
description of states with $M/J$ of order one. These are the
classical solutions of the Landau-Lifshitz sigma-model
\eqref{leadint}. It is clear that any finite size solution of the
sigma-model \eqref{leadint}, $i.e.$ solutions that extend out in a
finite area on the two-sphere, will correspond to an energy
$\tilde{H}$ of order $J$ since the action \eqref{leadint} is
proportional to $J$. Therefore we see that we can find semi-classical
solutions of the sigma-model in the decoupled theory. Note also that
it is clear from \eqref{leadint} that a finite-size solution on the
two-sphere can be well-described classically since the action
\eqref{leadint} will be large when $J$ is large.

In conclusion we have that semi-classical string states appears for
$\tilde{H} \sim J$, $i.e.$ for $\lambda/(E-J) \sim J$. Therefore, we
find semi-classical string states in planar $\CN=4$ SYM on $\R\times
S^3$ in the regime \eqref{regime}.

%%%%%%%%%%%%%%%%%%%%%%%%%%%%%%%%%%%%%%%%%%%%%%%%%%%%%%%%%%%%%
\section{Decoupling limit of strings on $\ads_5\times S^5$}
\label{sec:decstrings}

In this section we implement the decoupling limit \eqref{limit} on
type IIB string theory on $\ads_5\times S^5$. This is accomplished
by first considering the limit on a purely classical level. In this
way we obtain the Landau-Lifshitz model as the limiting sigma-model.
Subsequently we consider the quantum effects for the decoupling
limit. We show that the transverse modes decouple and we argue why
the quantum effects are under control in our limit \eqref{limit}
even though one naively seems to enter a quantum string regime.
Finally we argue that this means that we can match the leading
spectra \eqref{specgauge3} of gauge and string theory by taking the
limit \eqref{limit} on both sides of the AdS/CFT correspondence.
This furthermore includes the semi-classical states for which the
action is large on both the gauge theory and the string theory
sides.

%%%%%%%%%%%%%%%%%%%%%%%%%%%%%%%%%%%%%%%%%%%%%%%%%%%%%%%%%%%%%
\subsection{Classical limit of $\ads_5\times S^5$ sigma-model}
\label{sec:claslim}

In the following we take the limit \eqref{limit} of the classical sigma-model for $\ads_5\times S^5$. The quantum effects are considered in Section \ref{sec:quanlim}.

Note that the classical limit of the sigma-model considered in the following closely resembles the limit of Kruczenski in \cite{Kruczenski:2003gt}. However, even though these limits resemble each other on the level of the classical sigma-model, they are different for the quantum string theory since Kruczenski takes $J \rightarrow \infty$ keeping $\lambda/J^2$ fixed whereas we take $\lambda \rightarrow 0$ keeping $J$ fixed.

We are considering type IIB string theory on the $\ads_5\times S^5$
background with metric
\begin{equation}
\label{metads} ds^2 = R^2 \left[ - \cosh^2 \rho \, dt^2 + d\rho^2 +
\sinh^2 \rho \, (d\Omega_3')^2 + d\zeta^2 + \sin^2 \zeta \,
d\alpha^2 + \cos^2 \zeta \, (d\Omega_3)^2 \right]
\end{equation}
and the five-form Ramond-Ramond field strength
\begin{equation}
\label{fsads} F_{(5)} = 2 R^4 \left[ \cosh \rho \, \sinh^3 \rho \,
dt \, d\rho \, d\Omega_3' + \sin \zeta \, \cos^3\zeta \, d\zeta \,
d\alpha \, d\Omega_3 \right]
\end{equation}
We use in the following that
\begin{equation}
\label{R4} R^4 =  \lambda (\alpha')^2
\end{equation}
This relates the string parameters $R$ and $\alpha'$ to the 't Hooft
coupling $\lambda$ of $\CN=4$ SYM.
Using this the limit \eqref{limit} can be formulated in string theory variables as \eqref{strlimit}. However, we choose below instead to use the variables of the gauge theory.

We parameterize the three-sphere $\Omega_3$ as
\begin{equation}
\label{threesphere} (d\Omega_3)^2 = d\psi^2 + \cos^2 \psi d\phi_1^2
+ \sin^2 \psi d\phi_2^2 = d\psi^2 + d\phi_-^2 + d\phi_+^2 + 2 \cos
(2\psi) d\phi_- d\phi_+
\end{equation}
where $2 \phi_\pm = \phi_1 \pm \phi_2$. The energy $E$ and the
$SO(6)$ Cartan generators $J_i$, $i=1,2,3$, are given by
\begin{equation}
E = i \partial_t \spa J \equiv J_1 + J_2 = - i \partial_{\phi_+}
\spa S_z \equiv \frac{J_1 - J_2}{2} = - \frac{i}{2}
\partial_{\phi_-} \spa J_3 = - i
\partial_\alpha
\end{equation}
In the limit \eqref{limit} we only consider the charges $E$, $J_1$
and $J_2$. Together with the fact that on the gauge theory side we
decouple everything but the $SU(2)$ sector in the limit
\eqref{limit} it seems evident that we can work in the region
$\rho=\zeta =0$ of the $\ads_5\times S^5$ background
\eqref{metads}-\eqref{fsads}. In this region the background is
simply given by the metric $ds^2 = R^2 [ -dt^2 + (d\Omega_3)^2]$.
Thus, we take the bosonic sigma-model for $\R \times S^3$ to be the
starting point below. This is obviously only valid classically since
we can have quantum fluctions in the directions transverse to
$\rho=\zeta =0$ (along with fermionic fluctuations). We deal with
these issues in Section \ref{sec:quanlim}.

We take now as starting point the $\R \times S^3$ background $ds^2 =
R^2 [ -dt^2 + (d\Omega_3)^2]$ with $(d\Omega_3)^2$ given by
\eqref{threesphere}. Define
\begin{equation}
\label{thetaphi}
\theta \equiv 2 \psi - \frac{\pi}{2} \spa \varphi \equiv 2 \phi_-
\end{equation}
then we have the metric
\begin{equation}
\label{theRxS3}
ds^2 = R^2 \left[ - dt^2 + \frac{1}{4} (d\Omega_2)^2 + \left( d
\phi_+ + \frac{1}{2} \sin \theta d\varphi \right)^2 \right]
\end{equation}
with the two-sphere metric given as
\begin{equation}
(d\Omega_2)^2 =  d\theta^2 + \cos^2 \theta d\varphi^2
\end{equation}
To approach the right energy scale, we make the coordinate transformation
\begin{equation}
\label{tildet}
\tilde{t} = \frac{\lambda}{J^2} t \spa \chi = \phi_+ - t
\end{equation}
This ensures that $\tilde{H} \equiv (E-J)J^2/\lambda = i
\partial_{\tilde{t}}$ which precisely corresponds to the energy that we found was relevant in the sigma-model description on the gauge theory side \eqref{tildeH}. Moreover, we have that $J = - i \partial_\chi$ and
$S_z=-i\partial_\varphi$.
With this the metric \eqref{theRxS3} is
\begin{equation}
ds^2 = \sqrt{\lambda}\, \alpha' \left[ \frac{J^2}{\lambda}
d\tilde{t} \Big( 2 d\chi + \sin \theta d\varphi \Big)  + \frac{1}{4}
(d\Omega_2)^2 + \left( d \chi + \frac{1}{2} \sin \theta d\varphi
\right)^2 \right]
\end{equation}
Consider now the sigma-model Lagrangian
\begin{equation}
\label{lagr} \CL = - \frac{1}{2} G_{\mu \nu} h^{\alpha \beta}
\partial_\alpha x^{\mu} \partial_\beta x^{\nu}
\end{equation}
We pick the gauge
\begin{equation}
\label{gaugec} \tilde{t} = \kappa \tau \spa p_\chi = \mbox{const.}
\spa h_{\alpha\beta} = \eta_{\alpha\beta}
\end{equation}
with $p_\chi \equiv \partial \CL / \partial \partial_\tau \chi$.
Employing this, the Lagrangian \eqref{lagr} is found to be
\begin{eqnarray}
\CL &=& \frac{\sqrt{\lambda}\,\alpha'}{2} \left[ \kappa
\frac{J^2}{\lambda} \Big( 2 \partial_\tau \chi + \sin \theta
\partial_\tau \varphi \Big) + \frac{1}{4} \Big( (\partial_\tau
\theta)^2 + \cos^2 \theta (\partial_\tau \varphi)^2 - (\theta')^2 -
\cos^2 \theta (\varphi')^2 \Big) \right. \nn \\ && \left. + \Big(
\partial_\tau \chi + \frac{1}{2} \sin\theta \partial_\tau \varphi
\Big)^2 - \Big( \chi' + \frac{1}{2} \sin\theta \varphi' \Big)^2
\right]
\end{eqnarray}
The Virasoro constraints are $G_{\mu\nu} \partial_\tau x^\mu \partial_\sigma x^\nu = 0$ and $G_{\mu\nu} ( \partial_\tau x^\mu \partial_\tau x^\nu +
\partial_\sigma x^\mu \partial_\sigma x^\nu ) = 0$, giving
\begin{eqnarray}
0 &=& \sqrt{\lambda}\,\alpha' \left[ \kappa \frac{J^2}{\lambda}
\Big( \chi' + \frac{1}{2} \sin \theta \varphi' \Big) + \frac{1}{4}
\Big( \partial_\tau \theta \theta' + \cos \theta \partial_\tau
\varphi \varphi' \Big) \right. \nn \\ && \left. + \Big(
\partial_\tau \chi + \frac{1}{2} \sin\theta \partial_\tau \varphi
\Big) \Big( \chi' + \frac{1}{2} \sin\theta \varphi' \Big) \right]
\end{eqnarray}
\begin{eqnarray}
0 &=& \kappa \frac{J^2}{\lambda} \Big( 2 \partial_\tau \chi + \sin
\theta \partial_\tau \varphi  \Big) + \frac{1}{4} \Big(
(\partial_\tau \theta)^2 + \cos^2 \theta (\partial_\tau \varphi)^2 +
(\theta')^2 + \cos^2 \theta (\varphi')^2 \Big) \nn \\ &&  + \Big(
\partial_\tau \chi + \frac{1}{2} \sin\theta \partial_\tau \varphi
\Big)^2 + \Big( \chi' + \frac{1}{2} \sin\theta \varphi' \Big)^2
\end{eqnarray}
We record that
\begin{equation}
p_\chi \equiv \frac{1}{2\pi \alpha'} \frac{\partial \CL}{\partial
\partial_\tau \chi} = \frac{\kappa}{2\pi}
\frac{J^2}{\sqrt{\lambda}}+ \frac{\sqrt{\lambda}}{2\pi} \left(\partial_{\tau}\chi+
\frac{1}{2}\sin{\theta}\partial_{\tau}\phi\right)
\end{equation}
Note that the last term on the right-hand side goes away in the
$\lambda\to 0$ limit which means that the result for $p_{\chi}$ is
consistent with the above gauge choice \eqref{gaugec}. From this we
see furthermore that
\begin{equation}
J = \int_0^{2\pi} d\sigma p_\chi = \kappa
\frac{J^2}{\sqrt{\lambda}}
\end{equation}
This means that we have
\begin{equation}
\label{kappa} \kappa = \frac{\sqrt{\lambda}}{J}
\end{equation}

We take now the $\lambda \rightarrow 0$ limit of the Lagrangian and
the constraints. For the Lagrangian we get
\begin{equation}
\frac{1}{\kappa} \CL = \frac{\alpha' J}{2} \left[ \Big( 2 \dot{\chi}
+ \sin \theta \dot{\varphi} \Big) - \frac{1}{4} \Big( (\theta')^2 +
\cos^2 \theta (\varphi')^2 \Big) - \Big(\chi' + \frac{1}{2} \sin
\theta \varphi' \Big)^2 \right]
\end{equation}
where we defined the dot as a derivative with respect to
$\tilde{t}$. For the constraints we get
\begin{equation}
\label{chicon} \chi' = - \frac{1}{2} \sin \theta \varphi' \spa
\dot{\chi} = - \frac{1}{2} \sin \theta \dot{\varphi} - \frac{1}{8}
\Big( (\theta')^2 + \cos^2 \theta (\varphi')^2 \Big)
\end{equation}
We can now eliminate $\chi$ from the Lagrangian. Using the first
constraint we see that we only have a $\dot{\chi}$ term in the
Lagrangian without coupling to the other fields. Because of this, we
can ignore it in the Lagrangian, since omitting this term do not
affect the EOMs for the other fields. We can thus write the gauge
fixed Lagrangian
\begin{equation}
\frac{1}{\kappa} \CL_{\rm gf} = \frac{\alpha' J}{2}\left[ \sin \theta
\dot{\varphi} - \frac{1}{4} \Big( (\theta')^2 + \cos^2 \theta
(\varphi')^2 \Big) \right]
\end{equation}
This is then supplemented with the two constraints \eqref{chicon} that determine
$\chi$ from the other fields.
Writing the action for the gauge-fixed Lagrangian $\CL_{\rm gf}$ as
\begin{equation}
I = \frac{1}{2\pi\alpha'} \int d\tau \int_0^{2\pi} d\sigma \CL_{\rm gf} =
\frac{1}{2\pi\alpha'} \int d\tilde{t} \int_0^{2\pi} d\sigma
\frac{1}{\kappa} \CL_{\rm gf}
\end{equation}
we see that the action of the resulting effective sigma-model after
the $\lambda \rightarrow 0$ limit is
\begin{equation}
\label{thesigmod} I = \frac{J}{4\pi} \int d\tilde{t} \int_0^{2\pi}
d\sigma \left[ \sin \theta \dot{\varphi} - \frac{1}{4}\Big(
(\theta')^2 + \cos^2 \theta (\varphi')^2 \Big) \right]
\end{equation}
From the first constraint in \eqref{chicon} we see that the action
\eqref{thesigmod} should be supplemented by the condition that the
total world-sheet momentum is zero
\begin{equation}
\int_0^{2\pi} d\sigma \sin \theta \varphi' = 0
\end{equation}
We see that \eqref{thesigmod} precisely corresponds to the leading
order part \eqref{leadint} of the sigma-model action \eqref{contact}
derived on the gauge theory side. Thus, also on the string theory
side we regain the Landau-Lifshitz model. This is encouraging since
we are getting the same action on the gauge theory and string theory
sides of AdS/CFT by taking the same limit on both sides of the
correspondence. However on the string side our limit is taken, so
far, purely classically. This is, as we discuss below, also the
reason why we get the Landau-Lifshitz model exactly in
\eqref{thesigmod} whereas on the gauge theory side the leading order
action \eqref{contact} is only an approximation. This point will be
resolved in Section \ref{sec:quanlim}.

It is important to note that if we consider solutions of the
sigma-model \eqref{thesigmod} which are of finite-size on the
two-sphere then the action \eqref{thesigmod} is large when $J$ is
large. Therefore, already at this point we see that we can safely match
semi-classical states on the string theory side to semi-classical states
on the gauge theory side when $J$ is large. We comment further on
this below.

%%%%%%%%%%%%%%%%%%%%%%%%%%%%%%%%%%%%%%%%%%%%%%%%%%%%%%%%%%%%%
\subsection{Taking into account quantum effects}
\label{sec:quanlim}

In Section \ref{sec:claslim} we took the decoupling limit \eqref{limit} of type IIB string theory on $\ads_5\times S^5$ on the level of the classical sigma-model. In the following we consider the quantum effects to see how the limit \eqref{limit} works in the quantized string theory.

In Section \ref{sec:claslim} we saw that taking the limit \eqref{limit} classically gives the Landau-Lifshitz sigma-model \eqref{thesigmod} without assuming $J$ large. As noted above, this is a problem since on the gauge theory side the Landau-Lifshitz sigma-model is only valid for large $J$ \eqref{leadint}. This problem will be resolved in the following by taking into account the quantum effects.

In the AdS/CFT correspondence we have that planar $\CN=4$ SYM on $\R \times S^3$ is dual to tree-level string theory on the $\ads_5\times S^5$ background \eqref{metads}-\eqref{fsads}. Tree-level string theory means that we are considering first-quantized strings on $\ads_5\times S^5$. We can write schematically the full partition function for first-quantized type IIB strings on the $\ads_5\times S^5$ background
\eqref{metads}-\eqref{fsads} as
\begin{equation}
\label{stringpart} Z = \int [D h] [D x] [D S] e^{i I[h,x,S]}
\end{equation}
where $h$ is the world-sheet metric, $x$ the bosonic fields and $S$
the fermionic fields. We have from the background
\eqref{metads}-\eqref{fsads} that the action $I[h,x,S]$ is
proportional to $R^2/\alpha' = \sqrt{\lambda}$, assuming we keep
fixed the fields and the world-sheet metric. Therefore, it is customary to
regard $\sqrt{\lambda}$ as an effective string tension on the
$\ads_5\times S^5$ background.

Now, in the decoupling limit \eqref{limit} we take $\lambda
\rightarrow 0$ as part of the limit. It therefore looks like we
enter deep into the quantum string regime, since naively it seems
that $I[h,x,S] \rightarrow 0$. However, this is not the case. As can
be seen from the classical limit in Section \ref{sec:claslim}, the
modes with energies $E-J$ of order $\lambda$ in the limit
\eqref{limit} give a finite contribution to $I(h,x,S)$. This can be
seen from the fact that the modes for which the limiting action
\eqref{thesigmod} is finite also give a finite value to the full
action $I(h,x,S)$ in the $\lambda \rightarrow 0$ limit. Thus, for
these modes we can hope to have the quantum effects under control
even though $\lambda \rightarrow 0$. We should be careful however
because even though some modes give a finite contribution to
$I(h,x,S)$ there can be significant changes to the action due to
quantum effects.

\subsubsection*{Corrections to sigma-model action}

One possible source of change of the action is that in general the
target space background of a sigma-model receives $\alpha'$
corrections when imposing conformal invariance of the sigma-model.
If such corrections occur they could significantly change how the
sigma-model looks since, effectively speaking, we are taking
$\alpha' \rightarrow \infty$. However, the $\ads_5\times S^5$
background is known to be an exact background due to the large
amount of supersymmetry \cite{Kallosh:1998qs}, thus we can trust the
sigma-model in our limit. In fact, in
\cite{Metsaev:1998it,Kallosh:1998zx,Frolov:2006cc} exact gauge-fixed
Lagrangians for the Green-Schwarz superstring on the $\ads_5\times
S^5$ background are found.

\subsubsection*{Decoupling of transverse modes}

Another possible source of change of the action, which is more
difficult to address, are the modes that do not give a finite
contribution to $I(h,x,S)$. One set of such modes is the bosonic modes that correspond to fluctations transverse to $\rho=\zeta=0$. To understand these modes we rewrite the full metric \eqref{metads} for $\ads_5\times S^5$ as
\begin{equation}
\label{smartads} ds^2 = \cos^2 \zeta \, R^2 [ -dt^2 + (d\Omega_3)^2
] - R^2 (\sinh^2 \rho + \sin^2 \zeta) dt^2 + R^2 A_{ij} dx^i dx^j
\end{equation}
where $x^i$ are the remaining directions ($\rho$, $\zeta$, $\alpha$
and $\Omega_3'$) and $R^2 A_{ij}$ is the metric for these
directions. We see from this that setting $\rho=\zeta=0$ in \eqref{smartads} we end up with the metric on $\R \times S^3$ which is the starting point of our classical analysis in Section \ref{sec:claslim}.

Using now \eqref{smartads}, we can write the full bosonic sigma-model
Lagrangian for $\ads_5\times S^5$ as
\begin{equation}
\label{fullL} \frac{1}{\kappa} \CL_{\rm full} = \cos^2 \zeta
\frac{1}{\kappa} \CL + \frac{\sqrt{\lambda}}{2J} A_{ij} \dot{x}^i
\dot{x}^j - \frac{J}{2\sqrt{\lambda}} A_{ij} {x'}^i {x'}^j -
\frac{\alpha' J^3}{2 \lambda} ( \sinh^2 \rho + \sin^2 \zeta )
\end{equation}
We see that the term without derivatives in this Lagrangian
corresponds to the potential term
\begin{equation}
\label{potterm}
\frac{J^3}{4\pi \lambda} ( \sinh^2 \rho + \sin^2 \zeta )
\end{equation}
This is a confining potential. For $\lambda \rightarrow 0$ any mode with $\rho >0$ or $\zeta > 0$ will be driven towards
the origin $\rho=\zeta=0$ by the confining potential. I.e. if we
excite a mode so that $\rho > 0$ or $\zeta > 0$ then the energy of
such a mode would be proportional to $1/\lambda$ which means that
for $\lambda \rightarrow 0$ it would cost an infinite amount of
energy to make such an excitation. Therefore we get that only modes
with $\rho=\zeta=0$ survive the limit \eqref{limit}.

Notice that the decoupling of modes transverse to $\rho=\zeta=0$ is the string equivalent of the decoupling of the modes not in the $SU(2)$ sector on the gauge theory side (see Section \ref{sec:decgauge}). It is for instance evident that a mode with $J_3$ non-zero would get an infinite potential \eqref{potterm} of order $1/\lambda$ just as having a gauge theory state with non-zero $J_3$ also would be of order $1/\lambda$.

It would be interesting to add fermions to the Lagrangian
\eqref{fullL}. This can be done using the approaches of Metsaev and
Tseytlin \cite{Metsaev:1998it,Kallosh:1998zx} or Frolov et al \cite{Frolov:2006cc}. We expect that one can show the decoupling of the fermions in a similar way as the above argument for the bosonic directions.

\subsubsection*{Quantum effects from transverse modes and fermions}

As shown above all the modes which do not give a finite contribution have a
confining potential which freezes them and confines them to a point
in the $\lambda \rightarrow 0$ limit. However, they can still contribute through quantum corrections, as we now discuss.

One possible source of quantum corrections is from the zero-modes of the transverse modes and the fermions. We have shown above that near the point where the decoupled modes should be confined to, the modes have a harmonic oscillator potential.
Hence, the zero point energy could contribute. However, here we are
saved by the fact that we are close to a supersymmetric BPS state.
We are considering states with energies slightly above $E=J$, and
$E=J$ is a half BPS state. Therefore, the zero-point energy is
cancelled out by supersymmetry.

There is another possible source of quantum corrections. Since the transverse modes and the fermions become arbitrarily heavy in the decoupling limit \eqref{limit} we can integrate out these modes and obtain an effective sigma-model action for the surviving modes. Classically we have shown that we obtain the effective action \eqref{thesigmod} for the surviving modes. However, when taking quantum effects into account, in integrating out the transverse modes and fermions, the action \eqref{thesigmod} can receive corrections. This is possible because the decoupled modes can contribute when we go off-shell. We now consider how such corrections to the action \eqref{thesigmod} should appear.

We first remark that for $J$ large the states which have finite size
on the two-sphere parameterized by $\theta$ and $\varphi$ have a
large value for the action $I(h,x,S)$ as we can see from
\eqref{thesigmod}. Therefore, such string states are semi-classical
and the quantum corrections are suppressed. This means that in the
full effective action obtained by integrating out the decoupled
modes the part proportional to $J$ should be given by
\eqref{thesigmod}. From this we can conclude that we have matched
gauge theory and string theory, on the level of the sigma-model
model actions, for the part of the sigma-model action which is
proportional to $J$. This is one of the main results of this paper.
It means that we can reliably match semi-classical states with large
$J$ found from weakly coupled gauge theory in the limit
\eqref{limit} to semi-classical states found on the string side in
the same limit \eqref{limit}.

To go on, we should consider terms in the effective sigma-model action which go like powers of $1/J$, as compared to the leading part \eqref{thesigmod}. That $1/J$ is the effective expansion parameter is clear from the fact that $1/J$ is seen to be the effective $\alpha'$ in the leading action \eqref{thesigmod}. Moreover, one can show by considering string states on the pp-wave background considered in \cite{Bertolini:2002nr} that the next correction arises as a $1/J$ correction and the higher corrections furthermore come in powers $1/J^n$ \cite{stringpaper}. This is done by an analysis similar to the one of Callan et al \cite{Callan:2003xr,Callan:2004uv}. It is also clear from this analysis that no off-shell contribution from the decoupled modes can enter at order $1/J$. This is basically because computing the corrections to the surviving string states takes the form of quantum mechanical perturbation theory, with $1/J$ being the perturbation parameter. Since in quantum mechanical perturbation theory it is only at the second order that one can receive contributions from off-diagonal elements of the perturbation we can infer that it is only at order $1/J^2$ that we get off-shell contributions from the decoupled modes. From these considerations we can conclude that the first correction to the action \eqref{thesigmod} is of order $1/J^2$. We can write this as
\begin{equation}
\label{thesigmod2} I = \frac{J}{4\pi} \int d\tilde{t} \int_0^{2\pi}
d\sigma \left[ \sin \theta \dot{\varphi} - \frac{1}{4}\Big(
(\theta')^2 + \cos^2 \theta (\varphi')^2 \Big) + \frac{G[\theta,\varphi]}{J^2} + \CO(J^{-3})  \right]
\end{equation}
where $G[\theta,\varphi]$ is a function of $\theta$, $\varphi$ and their derivatives. We see from this that we can match gauge theory and string theory, on the level of the sigma-model action, for the part of the sigma-model action which is proportional to $J$ plus the part proportional to $J \cdot 1/J = 1$. Instead at order $J \cdot 1/J^2 = 1/J$ the decoupled modes can give new contributions to the effective action, as parametrized by $G[\theta,\varphi]$ in \eqref{thesigmod2}.

Comparing to the sigma-model action \eqref{contact} derived on the
gauge theory side from the ferromagnetic Heisenberg spin chain we
can now conjecture how the full effective action for the surviving
modes should look. At order $J^{1-2n}$ we get a contribution with
$2n$ derivatives with respect to $\sigma$ such that the full
effective action matches the action \eqref{contact} on the gauge
theory side. That integrating out the decoupled modes gives rise to
higher-derivative terms is natural. It is interesting to
note that each new derivative $\partial_\sigma$ comes with a $1/J$. This
could seem surprising since in \eqref{thesigmod} $1/J$ plays the
role of $\alpha'$. However, in the free limit \eqref{xy} of the
sigma-model it is not hard to check that the density of the
world-sheet momentum goes like $1/J$ which means that the operator
for the world-sheet momentum is proportional to $-(i/J)
\partial_\sigma$.

In conclusion we have found that to order $J \cdot 1/J^2 = 1/J$ the
sigma-model \eqref{thesigmod} is an accurate description for the
surviving modes of type IIB string theory on $\ads_5\times S^5$ in
the decoupling limit \eqref{limit}. This leading part of the
sigma-model action (see also \eqref{thesigmod2}) matches the leading
part of the sigma-model action \eqref{leadint} on the gauge theory
side, thus giving agreement between string theory and gauge theory
in the decoupling limit \eqref{limit} on both sides of the AdS/CFT correspondence, for the leading and first subleading order in an expansion in $1/J$.
Furthermore, we conjecture that the decoupled modes can be
integrated out on the string theory side to obtain the full
sigma-model action \eqref{contact} that is found on the gauge theory
side as an effective action for the surviving string modes.

Below in Section \ref{sec:matchspec} and further in Section
\ref{sec:semiclas} we examine the consequences of the above
results for the matching of the spectra of gauge theory and string
theory in the AdS/CFT correspondence.

%%%%%%%%%%%%%%%%%%%%%%%%%%%%%%%%%%%%%%%%%%%%%%%%%%%%%%%%%
\subsection{Matching of gauge and string theory spectra}
\label{sec:matchspec}

In the above we have found that we can match gauge theory and string
theory in terms of a sigma-model description in the decoupling limit
\eqref{limit} of the AdS/CFT correspondence for $J$ large. We now employ this to match the spectra for gauge theory and string theory in the limit
\eqref{limit}.

\subsubsection*{First-quantized string states}

On the gauge theory side, we have that the low energy spectrum of
$\tilde{H}$ for large $J$ is given by \eqref{specgauge3}. We now
argue that we can find the same spectrum on the string side in the
same regime. Consider the sigma-model action \eqref{thesigmod2}. As
for the sigma-model on the gauge theory side we zoom in near the point $(\theta,\varphi)=(0,0)$ on the two-sphere by taking the large $J$ limit with $\theta$ and $\varphi$ given by (see Eqs.~\eqref{xy} and \eqref{px})
\begin{equation}
x = \sqrt{J} \varphi \spa p_x = \frac{\sqrt{J}}{4\pi} \sin \theta
\end{equation}
This gives the Hamiltonian \eqref{Hcorr} that we obtained on the gauge theory side, valid up to $1/J^2$ corrections. Note here that $\tilde{H} = i \partial_{\tilde{t}}$ as defined above in Section \ref{sec:claslim}. Therefore, using the same procedure as in Section \ref{sec:smgauge} we obtain again the spectrum \eqref{specgauge3}. This means that up to order $1/J^2$, we find the same string spectrum from the decoupling limit \eqref{limit} of strings on $\ads_5 \times S^5$ as we found on the gauge theory side.

Note that the matching of the spectrum here is made in a $\lambda \rightarrow 0$ limit with $J$ large but finite. Therefore, in this sense we have that the string sigma-model action $I[h,x,S]$ in \eqref{stringpart} is finite after the decoupling limit.

We can thus conclude that our matching of the string theory sigma-model action \eqref{thesigmod2} to the leading order part of the gauge theory sigma-model action \eqref{contact} enables us to match the first-quantized string spectrum to the spectrum \eqref{specgauge3} found on the gauge theory side.

We emphasize that this matching of spectra is highly non-trivial in that on the string side we are taking a $R^2/\alpha' \rightarrow 0$ limit, which ordinarily would mean that the quantum corrections would become large. Instead, in the limit \eqref{limit} we have shown in Section \ref{sec:quanlim} that we can keep the quantum corrections under control by having $J$ large. A crucial part of this, shown in Section \ref{sec:claslim}, is that even though the string sigma-model action $I[h,x,S]$ is proportional to $R^2/\alpha'$ there is another part of the action multiplying this that diverges for $R^2 / \alpha' \rightarrow 0$ such that we end up with a finite action $I[h,x,S]$ in the \eqref{limit} limit.

Another related reason that the matching works is that on the gauge
theory side we are able to take a strong coupling limit even though
$\lambda \rightarrow 0$. This is due to the fact that the effective
coupling in the regime \eqref{regime} is not $\lambda$ but rather
$\lambda/(E-J)$. Therefore by having $\lambda/(E-J) \gg 1$ while
$\lambda \rightarrow 0$ we are accessing a strong coupling regime of
the gauge theory side, even though the 't Hooft coupling $\lambda$
is small. We discuss how to see that $\lambda/(E-J) \gg 1$ is a
strong-coupling regime in Section \ref{sec:hag} and in the
conclusion in Section \ref{sec:concl}.

\subsubsection*{Semi-classical string states}

As already anticipated in the end of Section \ref{sec:claslim}, we
can match the leading order contribution to the energy of a
semi-classical string state in the decoupling limit \eqref{limit} of
type IIB string theory on $\ads_5\times S^5$ (provided of course the
semi-classical string state survives the limit \eqref{limit}) to the
leading order energy of the corresponding state on the gauge theory
side. This follows from the matching of the string theory and gauge
theory sigma-model actions \eqref{thesigmod2} and \eqref{leadint} to
leading order for $J \rightarrow \infty$. That quantum corrections
cannot alter this result is due to the fact that $I[h,x,S]$ is of
order $J$ for a semi-classical string state.

Again, the matching of the classical energy of a string state to the
energy of a gauge theory state is rather non-trivial since we are
considering a $R^2/\alpha' \rightarrow 0$ limit in \eqref{limit}.
Thus the lesson here is that we can have a large string sigma-model
action $I[h,x,S]$ even though $\lambda \ll 1$. That this is possible
follows from the classical limit in Section \ref{sec:claslim} where
we saw that even though $R^2/\alpha' \rightarrow 0$ we still end up
with a finite action.

As a simple example of a semi-classical state we can consider the
rigid circular string solution \cite{Frolov:2003qc}
\begin{equation}
\theta = 0 \spa \varphi = 2 m \sigma
\end{equation}
This corresponds to having $\phi_1 = - \phi_2 = m \sigma$ and $\psi
= \pi /4$. The classical energy of this state is $\tilde{H} = J m^2
/ 2$.

Note that our results for the matching of the actions on the string
side \eqref{thesigmod2} and the gauge theory side \eqref{leadint}
also means that one can match $1/J$ corrections to the leading
classical result for the energy. This provides an explanation to the
many spectacular results for the matching of energies for
semi-classical states
\cite{Frolov:2003qc,Beisert:2005mq,Astolfi:2008yw}.

%%%%%%%%%%%%%%%%%%%%%%%%%%%%%%%%%%%%%%%%%%%%%%%%%%%%%%%%%%%%%
\section{Connection to semi-classical string regime}
\label{sec:semiclas}

In Section \ref{sec:decstrings} we have shown that we can match the
leading spectra of gauge and string theory by taking the limit
\eqref{limit} on both sides of the AdS/CFT correspondence. In this
section we argue that our results explains the matching between
weakly-coupled gauge theory and string theory in the semi-classical
regime at first order in $\lambda' = \lambda /J^2$.

The argument is rather simple. In Section \ref{sec:decgauge} and \ref{sec:decstrings} we have examined gauge theory and string theory in the regime
\begin{equation}
\label{gtreg}
\lambda \ll 1 \spa J \gg 1
\end{equation}
In this regime we can expand the energy/scaling dimension in $\lambda$ and $1/J$ on both the gauge theory and the string theory side. Expanding in $\lambda$ we can write the gauge theory energy/scaling dimension as
\begin{equation}
E_{\rm gt} - J = \lambda A_1 + \lambda^2 A_2 + \CO(\lambda^3)
\end{equation}
and the energy of string states as
\begin{equation}
E_{\rm str} - J = \lambda B_1 + \lambda^2 B_2 + \CO(\lambda^3)
\end{equation}
At each order in $\lambda$ we can then expand in $1/J$. At first order in $\lambda$ we have
\begin{equation}
A_1 = \frac{a_1}{J^2} + \frac{a_2}{J^3} + \CO(J^{-4})
\spa 
B_1 = \frac{b_1}{J^2} + \frac{b_2}{J^3} + \CO(J^{-4})
\end{equation}
Validity of the AdS/CFT correspondence in \eqref{gtreg} means that
$E_{\rm gt}=E_{\rm str}$. Indeed, we have shown in Section
\ref{sec:decstrings} that $a_1=b_1$ and $a_2= b_2$ in the $SU(2)$ sector. This result is thus a non-trivial confirmation of the AdS/CFT correspondence and it relied
on the result \eqref{thesigmod2} which shows that quantum
corrections to the classical sigma-model only enters at order
$1/J^2$ as compared to the leading term.

Consider instead the regime
\begin{equation}
\label{streg}
1 \ll \lambda \ll J^2
\end{equation}
This is a semi-classical regime for type IIB string theory since $\lambda \gg 1$.
In this regime we can expand the energy of string states in $\lambda' = \lambda/ J^2$
\footnote{Note that in this semi-classical regime there are also
non-analytical terms in $\lambda$ contributing
\cite{Beisert:2005cw,Hernandez:2006tk}. These terms comes from
quantum corrections to the classical string result and can be seen
as $1/\sqrt{\lambda}$ corrections to \eqref{exp69} and are therefore
small when $\lambda$ is large.}
\begin{equation}
\label{exp69} E_{\rm str} - J = \lambda' C_1 +
\lambda'^2 C_2 + \CO( {\lambda'}^{5/2} )
\end{equation}
At each order in $\lambda'$ we can then expand in powers of $1/J$
\begin{equation}
C_1 = c_1 + \frac{c_2}{J} + \CO( J^{-2} )
\end{equation}

Since the two regimes \eqref{gtreg} and
\eqref{streg} do not overlap there is a priori no reason why the
energies should agree in these two regimes. Indeed a mismatch can be resolved by introducing an interpolating function of $\lambda$ between the
two.%
\footnote{In the study of integrability of the AdS/CFT correspondence this has been achieved by the introduction of a phase-factor which changes as
one goes from $\lambda \ll 1$ to $\lambda \gg 1$
\cite{Beisert:2005tm,Beisert:2006ib,Beisert:2006ez}.}
However, it has been found in numerous computations, both for
quantum string states and semi-classical string states, that $a_1=c_1$ and $a_2 = c_2$
\cite{Minahan:2002ve,Beisert:2003tq,Beisert:2003xu,Callan:2003xr,Callan:2004uv,Frolov:2003qc,Beisert:2005mq,Astolfi:2008yw}.
This agreement begs for an explanation.

We can argue for  $a_1=c_1$ and $a_2 = c_2$ as
follows. In the regime \eqref{gtreg} we can infer from \eqref{thesigmod2} that the
computation of $b_1$ and $b_2$ only involves the classical
Landau-Lifshitz sigma-model. In the semi-classical regime \eqref{streg} we have the classical sigma-model limit for $\lambda \gg 1$. Therefore, to leading order we can compute the energy of string states from the classical Landau-Lifshitz sigma-model. This is in particular true for the $c_1$ and $c_2$ coefficients which are not affected by quantum corrections. Therefore, it follows that $b_1=c_1$ and $b_2=c_2$ since these coefficients are computed from the classical Landau-Lifshitz sigma-model in both regimes.
This agreement holds both for semi-classical string
states, with a large number of excitations, as well as for quantum string
states. Since $a_1=b_1$ and $a_2=b_2$ as consequence of our results in Section \ref{sec:decstrings}, we see that {\sl it
follows from our results that $a_1=c_1$ and $a_2=c_2$ in the $SU(2)$ sector}. Using our argument of Section \ref{sec:decstrings} we
have thus bridged the gap between the two non-overlapping regimes
\eqref{gtreg} and \eqref{streg} by connecting the two regimes on the
string theory side. We can therefore conclude that the agreement of weakly
coupled gauge theory and semi-classical string theory at one-loop up
to $1/J^2$ corrections ($i.e.$ $a_1=c_1$ and $a_2 = c_2$) is not a coincidence.

It has furthermore been found that $J^4 A_2 = C_2$ for the leading and first order correction in $1/J$
\cite{Beisert:2003tq,Callan:2003xr,Callan:2004uv}. It would be very
interesting if one could extend our arguments to get an
understanding of this agreement as well.

%%%%%%%%%%%%%%%%%%%%%%%%%%%%%%%%%%%%%%%%%%%%%%%%%%%%%%%%%%%%%
\section{The decoupling limit as a non-relativistic limit}
\label{sec:nonrel}

In this section we show that the decoupling limit \eqref{limit}
corresponds to a non-relativistic limit of type IIB string theory on
$\ads_5\times S^5$. We show this in full detail for type IIB string
theory on the maximally supersymmetric pp-wave in Section
\ref{sec:nonrelpp}. In Section \ref{sec:gennr} we extend the
analysis to the $\ads_5\times S^5$ background and we comment on the
relation to other non-relativistic limits in string and M-theory.

\subsection{Non-relativistic limit of string theory on pp-wave}
\label{sec:nonrelpp}

\subsubsection*{Penrose limit for flat-direction pp-wave}

We review here how to take the Penrose limit of \cite{Bertolini:2002nr}, giving the maximally supersymmetric pp-wave background of \cite{Blau:2001ne} in a coordinate system with a flat direction \cite{Michelson:2002wa,Bertolini:2002nr}. The $\ads_5 \times S^5$ background is \eqref{metads} and \eqref{fsads}. For simplicity, we ignore the five-form field strength and the fermionic fields in the following. Using the variables \eqref{thetaphi} along with
\begin{equation}
t' = t \spa \chi = \phi_+ - t
\end{equation}
and using \eqref{smartads}, we get the following metric for $\ads_5 \times S^5$
\begin{eqnarray}
ds^2 &=& R^2 \cos^2 \zeta \left[ 2 dt' d\chi + \sin \theta d\varphi
dt'  + \frac{1}{4} ( d\theta^2 + \cos^2\theta d\varphi^2) + \left( d
\chi + \frac{1}{2} \sin \theta d\varphi \right)^2 \right] \nn \\ &&
- R^2 ( \sinh^2 \rho + \sin^2 \zeta ) (dt')^2 + R^2 \left[  d\rho^2
+ \sinh^2 \rho  (d\Omega_3')^2 + d\zeta^2 + \sin^2 \zeta d\alpha^2
\right]
\end{eqnarray}
Here $E-J = i \partial_{t'}$, $J=-i\partial_\chi$ and $S_z = -i\partial_\varphi$.
Define now the coordinates $\gamma$, $x$, $y$, $r$ and $\tilde{r}$
by
\begin{equation}
\gamma = J \chi \spa x = \sqrt{J} \varphi \spa y =
\sqrt{J} \theta \spa r = \sqrt{J} \rho \spa \tilde{r}
= \sqrt{J} \zeta
\end{equation}
Then the Penrose limit is
\begin{equation}
\label{pplim} J \rightarrow \infty \spa \lambda' \equiv
\frac{\lambda}{J^2} \ \mbox{fixed}\spa \alpha' \ \mbox{fixed} \spa t',\, \gamma,\, x,\, y,\,
r,\, \tilde{r},\, \alpha, \, \Omega_3' \ \mbox{fixed}
\end{equation}
Taking the Penrose limit gives the metric
\begin{equation}
\label{ppmet} \frac{ds^2}{\alpha' \sqrt{\lambda'} } = 2 dt' d\gamma  +
\frac{1}{4} (dx^2+dy^2) + y dx dt'  + \sum_{i=1}^6 dz_i^2 -
\sum_{i=1}^6 z_i^2 (dt')^2
\end{equation}
Here the coordinates $z_1,...,z_4$ are defined by $r^2 =
\sum_{i=1}^4 z_i^2$ and $dr^2+r^2 (d\Omega_3')^2 = \sum_{i=1}^4
dz_i^2$ and $z_5$, $z_6$ are defined by $z_5+iz_6 = \tilde{r}
e^{i\alpha}$.
We see that this is the pp-wave background considered in \cite{Michelson:2002wa,Bertolini:2002nr}. Choosing the gauge
\begin{equation}
t' = c \tau \spa h_{\alpha\beta} = \eta_{\alpha\beta}
\end{equation}
we obtain the gauge fixed Lagrangian
\begin{equation}
\label{ppLgf}
\CL_{\rm gf} = \frac{c}{2} y \partial_\tau x + \frac{(\partial_\tau x)^2+(\partial_\tau y)^2}{8} - \frac{x'^2+y'^2}{8} + \frac{1}{2} \sum_{i=1}^6 \Big[ (\partial_\tau z_i)^2 - z_i'^2 - c^2 z_i^2 \Big]
\end{equation}
along with the action
\begin{equation}
I = \frac{\sqrt{\lambda'}}{2\pi} \int d\tau \int_0^{2\pi} d\sigma \CL_{\rm gf}
\end{equation}
From the term $c \partial_\tau \gamma$ in the full Lagrangian, the constant $c$ can be fixed to be
\begin{equation}
\label{cform}
c = \frac{1}{\sqrt{\lambda'}}
\end{equation}
The Hamiltonian is
\begin{equation}
H_{\rm lc} = \frac{\lambda'}{2\pi} \int_0^{2\pi} d\sigma \left\{ \frac{(\partial_\tau x)^2+(\partial_\tau y)^2}{8} + \frac{x'^2+y'^2}{8} + \frac{1}{2} \sum_{i=1}^6 \Big[ (\partial_\tau z_i)^2 + z_i'^2 +c^2 z_i^2 \Big] \right\}
\end{equation}
Defining $z(\tau,\sigma) = x(\tau,\sigma) + i y(\tau,\sigma)$, we can write the mode expansions of the bosonic fields as
\begin{equation}
\label{zmode}
z(\tau,\sigma) = 2 \sqrt{c} \, e^{i c \tau} \sum_{n \in \Z} \frac{1}{\sqrt{\omega_n}} \Big[ a_n e^{-i ( \omega_n \tau - n \sigma ) } -  \tilde{a}^\dagger_n e^{i ( \omega_n \tau - n \sigma ) } \Big]
\end{equation}
\begin{equation}
z_i (\tau,\sigma ) = i \frac{\sqrt{c}}{\sqrt{2}} \sum_{n\in \Z} \frac{1}{\sqrt{\omega_n}} \Big[ a^i_n e^{-i ( \omega_n \tau - n \sigma ) } - (a^i_n)^\dagger e^{i ( \omega_n \tau - n \sigma ) } \Big]
\end{equation}
where we used $\omega_n = \sqrt{n^2 + c^2}$. The canonical commutation relations $[x(\tau,\sigma),p_x(\tau,\sigma')] = i\delta (\sigma-\sigma')$, $[y(\tau,\sigma),p_y(\tau,\sigma')] = i\delta (\sigma-\sigma')$ and $[z_i(\tau,\sigma),p_j(\tau,\sigma')] = i\delta_{ij} \delta (\sigma-\sigma')$ follows from
\begin{equation}
\label{comrel}
[a_m,a_n^\dagger] = \delta_{mn} \spa [\tilde{a}_m,\tilde{a}_n^\dagger] = \delta_{mn} \spa [a^i_m,(a^j_n)^\dagger] = \delta_{mn} \delta_{ij}
\end{equation}
Employing \eqref{comrel} we obtain the bosonic spectrum
\begin{equation}
\label{Hlcspec}
c H_{\rm lc} = \sum_{n\neq 0} (\omega_n - c) M_n + \sum_{n\in \Z} (\omega_n + c ) N_n + \sum_{i=1}^6 \sum_{n\in \Z} \omega_n N^i_n
\end{equation}
with the number operators $M_n = a^\dagger_n a_n$, $N_n = \tilde{a}^\dagger_n \tilde{a}_n$ and $N^i_n = (a^i_n)^\dagger a^i_n$, and with the level-matching condition
\begin{equation}
\sum_{n\neq 0} n M_n + \sum_{n\in \Z} n N_n + \sum_{i=1}^6 \sum_{n\in \Z} n N^i_n = 0
\end{equation}

\subsubsection*{Decoupling limit as non-relativistic limit}

We now consider the decoupling limit \eqref{limit} and show explicitly that it is a non-relativistic limit.

As in \cite{Harmark:2006ta} we can take the limit directly of the spectrum \eqref{Hlcspec}. We first notice that the rescaled energy defined in \eqref{tildeH} is $\tilde{H} = c^2 H_{\rm lc}$. Since $c \rightarrow \infty$ we see that the modes with non-zero $N_n$ and $N^i_n$ become infinitely heavy, whereas the modes with $M_n$ gives the following spectrum
\begin{equation}
\label{galspec}
\tilde{H} = \frac{1}{2} \sum_{n\neq 0} n^2 M_n \spa \sum_{n\neq 0} n M_n = 0
\end{equation}
This match the leading part of \eqref{specgauge3}, in accordance with Sections \ref{sec:decstrings} and \ref{sec:semiclas} \cite{Harmark:2006ta}.

We see that the decoupling of the modes $N^i_n$ corresponds to the decoupling of the six modes transverse to $\R \times S^3$ as discussed in Section \ref{sec:quanlim}. Instead for the $N_n$ modes, we can interpret their decoupling as a consequence of the non-relativistic nature of the limit \eqref{limit}, as we now shall discuss.

The first hint that the limit \eqref{limit} is non-relativistic comes from considering the dispersion relation for the $M_n$ modes. Notice that a single mode with $M_n=1$ has
\begin{equation}
E - J = \sqrt{1 + \lambda' n^2 } - 1
\end{equation}
We can interpret this as a relativistic dispersion relation $\CE =
\sqrt{m^2+p^2}$ where the energy is $\CE = (E - J +
1)/\sqrt{\lambda'}$, the rest-mass is $m=1/\lambda'$ and the
momentum is $p=n$. The non-relativistic limit is then that $p/m
\rightarrow 0$ giving a Galilean dispersion relation $\CE - m = p^2
/ (2m) $. We see that this precisely is realized by the limit
\eqref{limit}.

Consider now the mode expansion \eqref{zmode} of the field $z(\tau,\sigma) = x(\tau,\sigma) + i y(\tau,\sigma)$. Before taking the limit we see that we have two sets of modes $a_{n \in \Z}$ and $\tilde{a}_{n \in \Z}$. This is in accordance with having two spatial directions. Considering the limit \eqref{limit} of \eqref{zmode}, we see that
\begin{align}
\label{limzmode}
z(\tau,\sigma) & = 2 \sum_{n \in \Z} \sqrt{ \frac{c}{\omega_n} } \Big[ a_n e^{-i (\omega_n - c) \tau + i n \sigma  } - \tilde{a}^\dagger_n e^{i  (\omega_n + c) \tau - i n \sigma  } \Big]
\nn \\ & \simeq 2 \sum_{n \in \Z} a_n e^{ - i \frac{n^2}{2} \tilde{t} + i n \sigma } - 2 \sum_{n \in \Z } \tilde{a}_n^\dagger e^{2 i c^2 \tilde{t} - i n \sigma}
\end{align}
where we used the rescaled time $\tilde{t} = \lambda' t' = \sqrt{\lambda'} \tau$, as introduced in Section \ref{sec:claslim}. The time $\tilde{t}$ measures the appropriate energy scale $\tilde{H}$ for the limit \eqref{limit}.
The first term in \eqref{limzmode} clearly becomes the mode expansion for the surviving modes after the limit, corresponding to the mode expansion \eqref{zz} for the free limit of the sigma-model describing the Heisenberg spin chain, see Section \ref{sec:smgauge}. The second term in \eqref{limzmode} decouples since $c^2 \rightarrow \infty$. Therefore, after the limit only the $a_{n \in \Z}$ modes are left. That half of the modes vanishes when going from a relativistic to a Galilean dispersion relation has a clear physical interpretation. This being that before the limit any particle mode has an anti-particle mode propagating backwards in time, as is the case in a relativistic field theory. After the limit we instead have Galilean symmetry, now with the anti-particle modes $\tilde{a}_{n \in \Z}$ decoupled. $I.e.$ the field $z(\tilde{t},\sigma)$ only contains lowering operators after the limit and has thus no anti-particle part.

We can furthermore see that the limit \eqref{limit} is
non-relativistic by considering the velocities. We have that $\tau =
c \tilde{t}$ and hence the velocities $\partial_\tau x$,
$\partial_\tau y$ and $\partial z_i$ all go to zero like $1/c$ as $c
\rightarrow \infty$. That the velocities go to zero is obviously a
clear signature of a non-relativistic limit. Taking the
$c\rightarrow \infty$ limit of the Lagrangian \eqref{ppLgf} we get
\begin{equation}
\CL_{\rm gf} = \frac{1}{2} y \partial_{\tilde{t}} x - \frac{x'^2+y'^2}{8} - \frac{1}{2} \sum_{i=1}^6 \Big[ z_i'^2 + c^2 z_i^2 \Big]
\end{equation}
We see that only $x(\tilde{t},\sigma)$ is dynamical after the limit.
The six transverse directions $z_i(\tau,\sigma)$ are non-dynamical
and decoupled, and the potential forces the string to be located at
$z_i=0$ which is also what we found in Section \ref{sec:quanlim}.
Taking the limit \eqref{limit} on the level of the action, we thus
get the action
\begin{equation}
\label{galact2}
I = \frac{1}{2\pi} \int d\tilde{t} \int_0^{4\pi} d\sigma \left[ y \partial_{\tilde{t}} x - \frac{x'^2+y'^2}{4} \right]
\end{equation}
This precisely correspond to the action \eqref{galact} obtained in the free limit of the sigma-model found on the gauge theory side. In Section \ref{sec:decvspp} we consider further the two ways of obtaining the action \eqref{galact2}.
As discussed in Section \ref{sec:smgauge} the action \eqref{galact2} is a theory with Galilean symmetry and spectrum \eqref{galspec}.

Consider the momenta
\begin{equation}
p_x = \frac{1}{4\pi} y + \frac{1}{8\pi c}\partial_\tau x \spa p_y = \frac{1}{8\pi c} \partial_\tau y
\end{equation}
Since $\partial_\tau x$ and $\partial_\tau y$ both go to zero as $1/c$ when $c\rightarrow \infty$, we have that $p_x \rightarrow y/(4\pi)$ and $p_y \rightarrow 0$ in the limit \eqref{limit}. From the canonical commutator  $[x(\tau,\sigma),p_x(\tau,\sigma')] = i\delta (\sigma-\sigma')$ we see now that
\begin{equation}
\label{xycom}
[x(\tau,\sigma),y(\tau,\sigma')] = 4\pi i\delta (\sigma-\sigma')
\end{equation}
Thus, $x$ and $y$ become non-commutative in the limit \eqref{limit}. Note here that before the limit we have that $[x(\tau,\sigma),y(\tau,\sigma')]=0$. However, the origin of this non-commutativity is the decoupling of the $\tilde{a}_{n\in \Z}$ modes.

The non-commutativity \eqref{xycom} connects also to another aspect
of the non-relativistic nature of the limit \eqref{limit}. As one
can see for example from \eqref{galspec} we have after the limit a
one-dimensional Galilean theory, instead of the two-dimensional
relativistic theory before the limit. Thus, we effectively go from
having two spatial directions $x$ and $y$, to having only one
spatial direction $x$ in the limit \eqref{limit}. Since $p_x
\rightarrow y/(4\pi)$ in the limit we see that the two spatial
directions become the directions in a two-dimensional phase space
for a single spatial direction. In this way we accomplish reducing
the dimension from two to one. This ties up with the
non-commutativity of $x$ and $y$ in \eqref{xycom} since now $y$ is
the momentum conjugate of $x$.

We discuss the relation to the literature on non-relativistic limits and string theories with Galilean symmetries below in Section \ref{sec:gennr}.

%%%%%%%%%%%%%%%%%%%%%%%%%%%%%%%%%%%%
\subsection{General considerations}
\label{sec:gennr}

In Section \ref{sec:nonrelpp} we have shown explicitly that the limit \eqref{limit} of type IIB string theory on the maximally supersymmetric pp-wave background \eqref{ppmet} is a non-relativistic limit. We now explain that the limit \eqref{limit} of type IIB string theory on $\ads_5\times S^5$ also corresponds to a non-relativistic limit. To see this consider the limit in Section \ref{sec:claslim}. From \eqref{gaugec} and \eqref{kappa} we see that for the fields $\varphi(\tau,\sigma)$ and $\theta(\tau,\sigma)$ the velocities $\partial_\tau \varphi$ and $\partial_\tau \theta$ go to zero since $\kappa\rightarrow 0$ in the limit \eqref{limit}. This is obviously a clear sign of taking a non-relativistic limit and is in close resemblance with the limit of the pp-wave considered in Section \ref{sec:nonrelpp}. If we consider the momenta conjugate to $\varphi$ and $\theta$ we have
\begin{equation}
p_\varphi = \frac{J}{4\pi} \sin \theta + \frac{\sqrt{\lambda}}{8\pi} \partial_\tau \varphi + \frac{\sqrt{\lambda}}{4\pi} \partial_\tau \chi \spa p_\theta = \frac{\sqrt{\lambda}}{8\pi} \partial_\tau \theta
\end{equation}
We get from this that $p_\varphi \rightarrow J \sin\theta /(4\pi)$ and $p_\theta \rightarrow 0$ in the limit \eqref{limit}. Using the same reasoning as in Section \ref{sec:nonrelpp} we see that we go from a theory with two spatial directions $\varphi$ and $\theta$ to a one-dimensional theory in which $\varphi$ and $\theta$ instead parameterizes the phase space. This is in accordance with the Landau-Lifshitz action \eqref{thesigmod}. Moreover, the above limit of the momenta $p_\varphi$ and $p_\theta$ shows that $\varphi$ and $\theta$ do not commute after the limit, in accordance with the analysis of Section \ref{sec:smgauge}. In conclusion, the limit \eqref{limit} is a non-relativistic limit of type IIB string theory on $\ads_5\times S^5$.

It is interesting to compare our non-relativistic limit of type IIB string theory on $\ads_5\times S^5$ to other non-relativistic limits of string and M-theory which have been considered previously. For a D-brane with a near-critical electric field it has been found that the open strings on the D-brane result in an open string theory with Galilean dynamics and space-time non-commutativity, while the closed string sector decouples \cite{Seiberg:2000ms,Gopakumar:2000na,Harmark:2000wv,Klebanov:2000pp,DeRisi:2002gt} and this has furthermore been generalized to other branes as well \cite{Gopakumar:2000ep,Bergshoeff:2000ai,Harmark:2000ff}. Building on these limits, it was found in \cite{Gomis:2000bd,Danielsson:2000gi,Danielsson:2000mu} that one can find non-relativistic closed string theories (NRCS's) with Galilean dynamics by taking a near-critical limit of string theory in the background of a near-critical field in which one of the directions compactified. A similar NRCS limit was furthermore found for type IIB string theory on $\ads_5\times S^5$ \cite{Gomis:2005pg}.

The features common between the NRCS limits and our limit
\eqref{limit} are that they are low energy limits, which in our case
corresponds to sending $E-J \rightarrow 0$, they are
non-relativistic limits, $i.e.$ limits of slow velocities and with
Galilean dynamics, and they have modes that become infinitely heavy
in the limit and therefore decouple. It seems on the other hand that
there is not any direct relation or duality between our limit
\eqref{limit} and the NRCS limits since we find a space-space
non-commutative target space and since in the NRCS theories a
compact direction is needed in order to take the limits (the
surviving closed strings all have non-zero winding). Moreover, we
find a truncation of the degrees of freedom of the theory in that
the dimension of the target space is reduced, in contrast with the
NRCS limits in which the dimension of the target space is preserved.

%%%%%%%%%%%%%%%%%%%%%%%%%%%%%%%%%%%%%%%%%%%%%%%%%%%%%%%%%%%%%
\section{Decoupling limit versus Penrose limit}
\label{sec:decvspp}

We consider in this section briefly the interplay between two
different kinds of limits that one can take of type IIB string
theory on $\ads_5\times S^5$. The first kind is the Penrose limit.
This is purely geometrical limit in which the number of degrees of
freedom is the same as before the limit, but the background of the
strings changes from $\ads_5\times S^5$ to the maximally
supersymmetric pp-wave background of \cite{Blau:2001ne}. The second
kind of limit is the decoupling limit \eqref{limit}, see also
\cite{Harmark:2007px} for other limits of this kind. This kind of
limit is not geometrical but is taken directly of the string theory,
as we did in Section \ref{sec:decstrings} for the $\ads_5\times S^5$
sigma-model. In this kind of limit we zoom in on a particular regime
of the theory in which some of the degrees of freedom decouple from
the spectrum due to the rescaling of the energy and moreover the
interactions between the surviving modes simplify.

\begin{figure}[ht]
\centerline{\epsfig{file=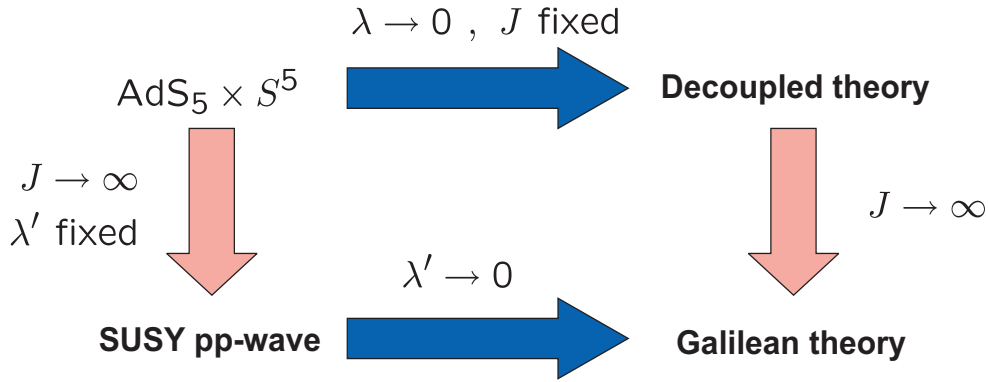,width=13cm,height=5cm} }
\caption{Overview of the limits. The limits going downwards corresponds to the Penrose limit and its manifestation for the decoupled theory. The limits going from left to right corresponds to the decoupling limit \eqref{limit} and its manifestation for the maximally supersymmetric pp-wave background.
\label{fig:limits}}
\end{figure}

In Figure \ref{fig:limits} we have illustrated the four limits that
we are considering. The top limit going from left to right is the
decoupling limit \eqref{limit} that we considered in Section
\ref{sec:decstrings}. The left limit going downwards is the Penrose
limit of $\ads_5 \times S^5$ \cite{Bertolini:2002nr} giving the
maximally supersymmetric pp-wave background of \cite{Blau:2001ne}.
We review briefly this limit in Section \ref{sec:nonrelpp}. Note
that the relevant Penrose limit is the one of
\cite{Bertolini:2002nr} rather than the one of
\cite{Blau:2002dy,Berenstein:2002jq}. We considered in Section
\ref{sec:nonrelpp} the limit on Figure \ref{fig:limits} on the
bottom going from left to right. This limit is the manifestation of
the decoupling limit for the maximally supersymmetric pp-wave
background and was found in \cite{Harmark:2006ta} (see also
\cite{Harmark:2006ie}). Finally, the right limit going downwards is
the manifestation of the pp-wave limit for the decoupled theory.
This limit is discussed in Section \ref{sec:smgauge} where it is
shown that we end up with a one-dimensional non-relativistic theory
with a Galilean symmetry, hence the name ``Galilean theory''.

As depicted in Figure \ref{fig:limits} one obtains the same
``Galilean theory'' irrespective of whether one first takes the
Penrose limit and then the decoupling limit or vice versa. Indeed this follows from the arguments of Sections \ref{sec:decstrings} and \ref{sec:semiclas}.
That the two limits in this sense commute, in that you end up with the same end point, is non-trivial since if we first take the pp-wave limit then $\lambda
\rightarrow \infty$ as part of that limit, whereas if we take the
other way around the diagram in Figure \ref{fig:limits} then we
always keep $\lambda$ small.

%%%%%%%%%%%%%%%%%%%%%%%%%%%%%%%%%%%%%%%%%%%%%%%%%%%%%%%%%%%%%
\section{Hagedorn temperature as interpolating quantity}
\label{sec:hag}

The limit \eqref{limit} was originally conceived as a limit of the
Grand Canonical partition function of $SU(N)$ $\CN=4$ SYM on $\R
\times S^3$. For our purposes we can write this partition function
as $Z(T,\Omega)$, where $T$ is the temperature and $\Omega$ is the
chemical potential associated to $J=J_1+J_2$. The limit in the Grand
Canonical ensemble takes the form
\cite{Harmark:2006di,Harmark:2006ta}
\begin{equation}
\label{grandlimit} \Omega \rightarrow 1 \spa \tilde{T} \equiv
\frac{T}{1-\Omega} \ \mbox{fixed} \spa \tilde{\lambda} \equiv
\frac{\lambda}{1-\Omega} \ \mbox{fixed} \spa N \ \mbox{fixed}
\end{equation}
The resulting partition function is
\cite{Harmark:2006di,Harmark:2006ta}
\begin{equation}
Z( \tilde{\beta} ) = \tr \Big[ e^{- \tilde{\beta} ( D_0 +
\tilde{\lambda} D_2 ) } \Big]
\end{equation}
where the trace is only over the $SU(2)$ sector and $\tilde{\beta} =
1/\tilde{T}$. In the planar limit $N=\infty$ one finds
\cite{Harmark:2006di,Harmark:2006ta}
\begin{equation}
\log Z (\tilde{\beta}) = \sum_{n=1}^\infty \sum_{J=1} \frac{1}{n}
e^{-\tilde{\beta} n L} Z_J^{\rm (XXX)} (n \tilde{\beta} )
\end{equation}
where $Z_J^{\rm (XXX)} (\tilde{\beta} )$ is the partition function
for the ferromagnetic Heisenberg spin chain of length $J$ with
Hamiltonian $\tilde{\lambda} D_2$. Thus, as in the microcanical
ensemble, planar $\CN=4$ SYM on $\R \times S^3$ in the limit
\eqref{grandlimit} is given exactly by the Heisenberg spin chain.
Using this it was found in \cite{Harmark:2006ta} that the Hagedorn
temperature is determined from the thermodynamic limit of the
free-energy per site of the ferromagnetic Heisenberg spin chain
\begin{equation}
f(t) = - t \lim_{J \rightarrow \infty} \frac{1}{J} \log \left[
{\tr}_J \left( e^{-t^{-1} D_2 } \right) \right]
\end{equation}
from the formula
\begin{equation}
\label{genhag} f\big( (\tilde{\beta}_H \tilde{\lambda} )^{-1} \big)
= - \tilde{\lambda}^{-1}
\end{equation}
Using \eqref{genhag} one can find the Hagedorn temperature
$\tilde{T}_H (\tilde{\lambda})$ as function of $\tilde{\lambda}$.
This was done in \cite{Harmark:2006ta} both for $\tilde{\lambda} \ll
1$ and $\tilde{\lambda} \gg 1$. More generally, we can infer from
\eqref{genhag} that we can interpolate the Hagedorn temperature
$\tilde{T}_H$ from $\tilde{\lambda} \ll 1$ to $\tilde{\lambda} \gg
1$.

For $\tilde{\lambda} \ll 1$ we can connect to the loop corrections
in weakly coupled $\CN=4$ SYM. More specifically, for
$\tilde{\lambda} \ll 1$ each term of power $\tilde{\lambda}^k$ in
$\tilde{T}_H$  origins from a $k$-loop correction in weakly coupled
$\CN=4$ SYM \cite{Harmark:2006ta}. Therefore, if we instead consider
$\tilde{\lambda} \gg 1$ we can infer that this is a strong-coupling regime
of $\CN=4$ SYM, even though we have $\lambda \ll 1$. Having $\tilde{\lambda} \gg 1$  is equivalent to having $\lambda/(E-J) \gg 1$ in the microcanonical ensemble.%
\footnote{To see this in detail, one uses that $1-\Omega$ sets the
energy scale in the limit \eqref{grandlimit}, $i.e.$ $\beta (E-J) =
\tilde{\beta} (E-J)/(1-\Omega)$, so states which contribute to the
partition function has $E-J \lesssim 1-\Omega$. Hence
$\tilde{\lambda} = \lambda /(1-\Omega) \gg 1$ is translated in the
microcanonical ensemble to $\lambda /(E-J) \gg 1$.} Thus we can
conclude that $\lambda/(E-J) \gg 1$ corresponds to a strong coupling
regime of $\CN=4$ SYM, even though $\lambda \ll 1$.

For $\tilde{\lambda}\gg 1$ it was found that $\tilde{T}_H =
(2\pi)^{1/3} \zeta(3/2)^{-2/3} \tilde{\lambda}^{1/3}$
\cite{Harmark:2006ta}. This result is obtained from the spectrum
\eqref{string0}. Since we have shown in this paper that this
spectrum can be found both from the gauge theory side as well as the
string theory side of AdS/CFT in the limit \eqref{limit} we can
match this Hagedorn temperature to the Hagedorn temperature of type IIB
string theory on $\ads_5\times S^5$. This was done in
\cite{Harmark:2006ta}. However, here we justify the steps of
\cite{Harmark:2006ta} in which the Hagedorn temperature was found on
the string theory side by first taking the Penrose limit
\eqref{pplim} and subsequently taking the limit \eqref{limit}.
Indeed, we have shown in Sections
\ref{sec:decstrings}-\ref{sec:decvspp} (see in particular the
commuting diagram in Figure \ref{fig:limits}, Section
\ref{sec:decvspp}) that one obtains the same spectrum
\eqref{string0} by first taking the limit \eqref{limit} and
subsequently considering $J\rightarrow \infty$. Therefore, we can
conclude that the Hagedorn temperature constitutes the first example
of a quantity, not protected by supersymmetry, which we can
interpolate fully from weak to strong coupling in AdS/CFT.

%%%%%%%%%%%%%%%%%%%%%%%%%%%%%%%%%%%%%%%%%%%%%%%%%%%%%%%%%%%%%
\section{Conclusions}
\label{sec:concl}

The basic idea of this paper is that we can compare gauge theory and
string theory quantitatively in the regime \eqref{regime} of the
AdS/CFT correspondence. The special thing about the regime
\eqref{regime} is that the 't Hooft coupling is small which means we
can compute the spectrum of states exactly using weakly coupled
$\CN=4$ SYM. Ordinarily, this would mean that we are deep in a
quantum string regime on the string theory side, since the length
scale of the $\ads_5\times S^5$ geometry is much smaller than the
string length, but we show in this paper that we can find a
semi-classical string theory regime as part of the regime
\eqref{regime}. This is related to the fact that while $\lambda \ll
1$ we have that $\lambda/(E-J) \gg 1$ which effectively means that
we are in a strong-coupling regime of $\CN=4$ SYM.

That $\lambda/(E-J) \gg 1$ is a strong-coupling regime of $\CN=4$
SYM is tied to the fact that we can see strings with continuous
world-sheets in the regime \eqref{regime}. In particular, we find in
this paper that there are semi-classical strings with $\lambda/(E-J)
\sim J$ and that single quantum strings which are weakly interacting
on the world-sheet appear for $\lambda/(E-J) \sim J^2$. As we
explain in this paper, it is the ability to match the energies for
such strings that makes a quantitative match of the spectra of
planar $\CN=4$ SYM on $\R\times S^3$ and type IIB string theory on
$\ads_5\times S^5$ in the regime \eqref{regime} possible.

We explore the regime \eqref{regime} by taking the decoupling limit
\eqref{limit}. We show that this limit can be seen as a
non-relativistic of strings on $\ads_5\times S^5$, and it is
conjectured to give a consistent truncation of both the gauge theory
and string theory sides of the AdS/CFT correspondence. For planar
$\CN=4$ SYM and type IIB string theory with $g_s=0$ we show that one
can match the leading terms in a sigma-model action in the
decoupling limit \eqref{limit} and for large $J$. This relies on our
identification of a new semi-classical regime for the string side.
Employing this, we match the spectra up to $1/J^2$ corrections. We
explain that this result shows why it has been found in several ways
that the one-loop contribution to the spectrum matches the string
theory spectrum up to $1/J^2$ corrections.

The results of this paper give a better understanding of the
matching of Hagedorn temperature in \cite{Harmark:2006ta}. We
conclude that the Hagedorn temperature constitutes the first example
of a quantity, not protected by supersymmetry, which we can
interpolate fully from weak to strong coupling in AdS/CFT.

Given that we have managed to identify a regime of the AdS/CFT
correspondence in which we can quantitatively match gauge theory and
string theory, it is interesting to ask what future applications
this can have. A very interesting direction taken in
\cite{stringpaper}%
\footnote{See also \cite{Astolfi:2008yw}.} is to match the spectrum
of type IIB strings on $\ads_5\times S^5$ to the spectrum obtained
on the gauge theory side at order $1/J^2$ in the limit
\eqref{limit}. The matching at order $1/J^2$ would be highly
interesting in that it should involve a non-trivial contribution
coming from integrating out the modes that decouple in the limit
\eqref{limit}, giving rise to a higher-derivative term in the
effective sigma-model description of the strings.

Another direction that one could pursue is to compute $\lambda$
corrections on the string theory side. This seems challenging on the
string theory side since one in principle should integrate out the
heavy decoupled modes order by order in $\lambda$ and then do
quantum mechanical perturbation theory in $\lambda$. However, if one
succeeds it could provide an alternative and more direct path to
resolving the famous three-loop discrepancy
\cite{Callan:2003xr,Callan:2004uv,Serban:2004jf}.

Another very interesting avenue to explore is to move away from the
planar limit and $g_s=0$. One direction could be to take a new look
at $1/N$ corrections.%
\footnote{See
\cite{Dobashi:2004nm,Grignani:2005yv,Grignani:2006en,Grignani:2007zz,Casteill:2007td}
for recent developments in computing $1/N$ corrections to string
states.} It is conceivable that the fact that string theory
simplifies in the limit \eqref{limit} could help in going further in
this direction. Another interesting direction would be to explore
the regime of finite
string coupling where one should see black holes.%
\footnote{See
\cite{AlvarezGaume:2006jg,Hollowood:2006xb,Silva:2006st,Dey:2007vt,Balasubramanian:2007bs,Hollowood:2008gp,Murata:2008bg}
for recent work on connecting the thermodynamics of $\CN=4$ SYM on
$\R \times S^3$ with the thermodynamics of black holes in
$\ads_5\times S^5$. Note in particular \cite{Hollowood:2008gp} in
which they consider a region with near-critical chemical potential.}
Obviously, if one could use the regime \eqref{regime} and the
decoupling limit \eqref{limit} to quantitatively match gauge theory
with black holes on the string theory side, it would be a result of
tremendous importance.

Finally, it is interesting to generalize our results to gauge/string
dualities with less supersymmetry. We note in particular the papers
\cite{Hikida:2006qb,Grignani:2007xz,Larsen:2007bm,Hamilton:2007he}
in which the thermodynamics are considered for gauge/gravity
correspondences with less than maximal supersymmetry. We remark
furthermore that we found a decoupling limit similar to
\eqref{limit} for pure Yang-Mills theory in \cite{Harmark:2007px}.

%%%%%%%%%%%%%%%%%%%%%%%%%%%%%%%%%%%%%%%%%
\section*{Acknowledgments}

We thank Gianluca Grignani and Shinji Hirano for useful discussions. TH and MO thank the Carlsberg foundation for support.

%The following two lines is for bibtex only:
%\bibliographystyle{C:/BIB/utphys}
%\bibliography{C:/BIB/mybib,C:/BIB/bibrot,C:/BIB/biblioniels}
%\bibliographystyle{../INPUT/newutphys_notitle}
%\bibliography{../BIB/mybib,../BIB/bibrot,../BIB/biblioniels}

\providecommand{\href}[2]{#2}\begingroup\raggedright\endgroup

\end{document}